\documentclass{COMNET}

\usepackage{amsmath} 
\usepackage{amsthm} 
\usepackage{hyperref} 
\usepackage{graphics} 
\usepackage{url} 

\usepackage{booktabs}
\usepackage{subcaption}
\usepackage{graphicx}

\begin{document}
	
	\title{Popularity and Centrality in Spotify Networks: Critical transitions in eigenvector centrality}
	
	\shorttitle{Popularity and Centrality} 
	\shortauthorlist{T. South, M. Roughan, L. Mitchell} 
	
	\author{
		\name{Tobin South$^{1,2,*}$, Matthew Roughan$^{1,2}$ and Lewis Mitchell$^{1,2}$}%
		\address{$^1$School of Mathematical Sciences, University of Adelaide, Adelaide, Australia.  \\ $^2$ARC Centre of Excellence for Mathematical \& Statistical Frontiers (ACEMS).\email{$^*$Corresponding author: tobin.south@adelaide.edu.au}}
	}


%
%


\begin{abstract} 
{
	The modern age of digital music access has increased the availability of data about music consumption and creation, facilitating the large-scale analysis of the complex networks that connect musical works and artists. Data about user streaming behaviour, and the musical collaboration networks are particularly important with new data-driven recommendation systems. Here we present a new collaboration network of artists from the online music streaming service Spotify, and demonstrate a critical change in the eigenvector centrality of artists, as low popularity artists are removed. This critical change in centrality, from a central core of classical artists to a core of rap artists, demonstrates deeper structural properties of the network. Both the popularity and degree of collaborators play an important role in the centrality of these groups. Rap artists have dense collaborations with other popular artists whereas classical artists are diversely connected to a large number of low and medium popularity artists throughout the graph through renditions and compilations. A Social Group Centrality model is presented to simulate this critical transition behaviour, and switching between dominant eigenvectors is observed. By contrasting a group of high-degree diversely connected community leaders to a group of celebrities which only connect to high popularity nodes, this model presents a novel investigation into the effect of popularity bias on how centrality and importance are measured.%
}
{collaboration networks, centrality, critical transitions, social network analysis}
\end{abstract}



\maketitle


\section{Introduction}

Social network analysis has often been a topic of scientific
interest in music~\cite{park_topology_2015, gleiser_community_2003, park_social_2007, crossley_pretty_2008, mcandrew_music_2015, bae_multi-scale_2016}, but as online streaming services become ubiquitous, there is an increasing need for them to monetise their operations, often through targeted advertising. There is also an increasing value in providing a good recommender system to keep customers (subscription or otherwise) engaged in the platform. Network analysis can provide valuable tools to help in these goals.

Network centrality metrics quantify the influence or importance of a node in a network.  Eigenvector centrality~\cite{bonacich_factoring_1972} and its variants are one of
the most widely used~\cite{wasserman_social_1994} because of their strong mathematical foundations, intuitive meaning, and also, importantly, because they are considered to be robust metrics.  Most network measurements suffer from various types of noise (sampling bias, missing data, aliasing, to name a few) and so it is critical to have a metric that is  insensitive to measurement noise.

The music industry has long used a variety of metrics to quantify the success of the music (e.g sales numbers, top 10 lists), but the new technology-driven industry of music streaming presents new demands to analyse music for purposes of ranking and recommendation. Collaboration graphs can provide a useful source of data as seen through the analysis of mentorship in electronic music~\cite{janosov_elites_2020} and the collaboration of movie actors~\cite{albert_statistical_2002,barabasi_emergence_1999}. The use of networks and complexity more broadly are powerful tools in the analysis of art~\cite{perc_beauty_2020,sigaki_history_2018,klimek_fashion_2019}.

This quantitative network approach to understanding music has been facilitated by the rise of large-scale datasets of music and the near global access to online music consumption technologies.
Accompanying the de-localisation of communities by the
Internet~\cite{dimaggio_social_2001}, musical collaboration scenes have become global~\cite{youngblood_cultural_2019}.  With this online communities can play an important role in music
discovery~\cite{garg_measuring_2011} and allow artists to establish collaborations and relationships regardless of location or distance~\cite{kruse_local_2010}, creating a global collaboration network for musical artists.
The rise of large music streaming platforms and associated data sources presents a tremendous yet underutilised opportunity for data mining and analysis. While work has been done studying musical sampling and collaboration networks using these sources~\cite{park_topology_2015, bae_multi-scale_2016, youngblood_cultural_2019},  the combination of social network analysis combined with music consumption data provides an opportunity to examine network properties at scales ranging from the superstars of the modern music industry to large diverse connected networks of artists big and small.

In this work we show that despite common intuition that it is a robust metric, eigencentrality has a surprising form of
sensitivity. We explore a large dataset comprising the
inter-collaboration graph of musical artists exposed by the online music streaming platform Spotify. The data we collected from Spotify contains over 1 million artists, along with information on which other artists collaborated with them (including sampling and other direct influences even when one artist is deceased). The network thus formed can reveal much about which artists are most influential.

Additionally, Spotify provides up-to-date data on the popularity of each artist. This provides another (non-network) based view of the importance of an artist. The two are somewhat correlated, but not completely. For instance, classical artists (e.g., Mozart, Bach) are often very influential, but more modern artists (e.g., Lil Wayne, Rick Ross, 2 Chainz) are more popular. Both views are important: the popularity metric provides an easily understandable view of importance, but if used exclusively in recommendations would lead to a ``rich get richer'' phenomenon which would squeeze out the many smaller artists on such platforms, whose existence is important to create a vibrant ecosystem. 

The interesting result arises when one metric is used to explore the other. In the case in point, we imagine that someone studying the network samples more popular artists first. In doing so, they should see important parts of the network without the computational and communications load of sampling all million+ artists and their collaboration history.  However, our results show that under such a sampling strategy, eigencentrality undergoes a critical transition where suddenly the group of most central artists swaps {\em en masse} for another group.

The reason this occurs is that eigencentrality uses the values in the eigenvector corresponding to the largest eigenvalue as its measure of centrality. This has multiple intuitive meanings: for instance, mathematically it can be seen as being proportional to the stationary probabilities of being at each node in a random walk across the network (similar to PageRank).  However, as the network is progressively changed, the eigenvalues change, and at some point the first and second swap dominance, resulting in a sudden shift to the erstwhile second eigenvector as the measure of centrality. 

Although this arises in the context of a particular dataset and sampling strategy, this problem is not linked solely to this situation. Any strategy that progressively changes a network by increasingly sampling it, or observing it as it evolves over time could suffer from this problem. 

To study this, we present a model of social group connection. This model is creating by considering two alternating views of impact in networks; grass-roots impact and popularity-dominant influence. The two groups formed from these view, ``community leaders'' and ``celebrities'' are seen to undergo the same critical transitions seen in the Spotify data. 

Summarising, the main contributions of this paper are:
\begin{itemize}
  \item the analysis of a network dataset from Spotify, including over 1 million artists;
  
  \item the discovery of a phase transition in eigencentrality metrics on samples of graphs; and 
  
  \item a network model of social group connection, through which these phase transitions in centrality can be studied.
  
\end{itemize}

\section{Background}
\subsection{Network Centrality}
Network structures are diverse and heterogeneous, as are the centrality measures on them. It is generally accepted that some nodes in a network can be considered as more important than other ones; however, there is no generalised definition of centrality in networks and, as a result, many centrality measures exist~\cite{costa_characterization_2007}. Here we list several common centrality measures as well as their interpretations and shortcomings.

\begin{itemize}
  \item The most simple centrality measure is the \emph{degree centrality}, which is defined by the number of edges attached to a node. Defining this in terms of the network's adjacency matrix $A$ with $N$ nodes gives,
  \begin{equation}
  {\mathbf c}_{i}^{\rm(degree)}=\sum_{j=1}^{N} A_{i j}.
  \end{equation}
  This centrality, although appealingly simple, has many drawbacks. For instance, nodes with high degree may exist on the periphery of networks, and in many human-made networks this metric can be gamed. 
  
  \item Centrality can also be obtained from the network's shortest paths. Defined using the length of the shortest path between nodes $i$ and $j$, $d_{ij}$, the \emph{closeness centrality} can be calculated using,
  \begin{equation}
  {\mathbf c}_{i}^{\rm (closeness)}=\frac{N}{\sum_{j=1, j \neq i}^{N} d_{i j}}.
  \end{equation}
  This centrality, although appealing due to its intuitive notion of distance, requires the calculation of shortest paths, which is notoriously computationally expensive for large graphs such as the one we will consider.
  
  \item To that note, \emph{betweenness centrality} is similarly built from shortest paths, using the number of shortest paths from nodes $a$ to $b$, $\sigma_{ab}$, compared to the number of those shortest paths that pass through a node $i$, $\sigma_{ab}(i)$. Defined by,
  \begin{equation}
  {\mathbf c}_{i}^{\rm(betweenness)}=\sum_{a \neq i \neq b} \frac{\sigma_{ab}(i)}{\sigma_{ab}},
  \end{equation}
  this metric can prove costly to compute on large graphs.
  
  \item  We can also define centrality recursively in terms of the centrality of a node's neighbourhood. 
  This is built from the notion that a node is important if it is connected to other important nodes.
  If this importance is given by the vector $v$, it can be defined through an eigenvector equation.
  Thus, let the importance of each node be a corresponding element in the vector $v^{\rm (eig)}$, which then gives the definition,
  \begin{equation}\label{eq:eigenvector}
  v_{i}^{\rm(eig)}=\frac{1}{\lambda} \sum_{k=1}^{N} A_{k, i} v_{k},
  \end{equation}
  with a constant $\lambda \neq 0$. In matrix form, this can be restated as,
  \begin{equation}
  \lambda v^{\rm(eig)} = A v^{\rm(eig)}, 
  \end{equation}
  which can be solved as the dominant left-hand eigenvector of the adjacency matrix $A$. 
  
  This has an intuitive interpretation as being proportional to the number of visits to each node on a random walk of infinite length~\cite{rodrigues_network_2019}. Further, this eigenvector can be numerically computed, even on large networks, using power iteration, 
  \begin{equation}
  v^{(k)}=v^{(k-1)} A.
  \end{equation}
  
  Although \emph{eigenvector centrality} (\emph{eigencentrality}) has some flaws, such as undesirable
  localisation of centrality in the
  network~\cite{martin_localization_2014}, it still remains a useful
  tool when dealing with large networks. 
  Indeed, eigencentrality has been shown to be more robust to conditions of imperfect data~\cite{costenbader_stability_2003} and network manipulation \cite{niu_robustness_2015} than other centrality measures.
  
  \item Similar to eigencentrality, \emph{PageRank}~\cite{page_pagerank_1999} is based on random walks. The differences are to transform the adjacency matrix such that the elements represent the probability of transition between nodes, and to allow for the walk to become ergodic (in the case of disconnected networks or only-weakly connected directed graphs), by introducing random restarts through a damping factor, $d$. Mathematically,
  \begin{equation}
  {\mathbf c}_i^{\rm (PageRank)}=\frac{1-d}{N}+d \sum_{j \in M(i)} \frac{{\mathbf c}_j^{\rm (PageRank)}}{\sum_k A_{jk}},
  \end{equation}
  where $M(i)$ is the set of nodes connected to node $i$. This formulation is incredibly useful for weakly-connected directed networks, such as the hyper-link web graph.
\end{itemize}

Notably, both eigencentrality and PageRank can be calculated using power iteration. However, the rate of convergence can reduce significantly if the difference between the first and second largest eigenvalues is small~\cite{haveliwala_second_2003}; a fact that will effect our calculations in this network.

Herein, we focus on eigencentrality for its intuitive interpretation, simple construction, and computational efficiency in our large Spotify collaboration network.   

\section{Data Collection and the Network}
\subsection{Spotify}
Spotify\footnote{\url{https://www.spotify.com}} is a digital music streaming service with over 50 million music tracks~\cite{noauthor_spotify_nodate}. These music tracks and metadata can be accessed through the Spotify Web Application Programming Interface (API)\footnote{\url{https://developer.spotify.com/documentation/web-api/}}. An artist's entire discography on Spotify can be accessed via this API with metadata about what other artists `performed' on the track as well as metadata about the artist such as their number of in-platform followers, a list of genres associated with the artist according to Spotify's internal analysis (many artists have not been classified), and a measure of the popularity of an artist. The popularity of an artist is an integer from 0 to 100, with 100 being the most popular artist on the platform. The artist popularity is calculated from the total number of plays of all the artist’s tracks and how recent those plays are.

API calls were made to collect the collaboration data for each artist's discography, starting from Kayne West (chosen for a high likelihood of being in the largest
connected component due to his high connectivity and popularity). All
artists that appeared as having `performed' on any track in an artists
discography were added to the search queue and these artists were
considered to have `collaborated' with the searched artist. This
snowball search of the artists using the API was continued for over a
month in December 2017 until the queue was exhausted in January 2018. In total
1,250,065 artists were found and extra metadata such as popularity and
genre were collected for a smaller subsample of 918,504 artist due to
API rate limits during the collection time frame. The size of this
subsample cannot meaningfully be increased without recollecting the
whole network due to changes in artist metadata since the collection
period.  We chose to collect as large a subsample as we could over the predefined study period, and this 73\% sample of the population is sufficient to calculate statistics robustly.

From this snowball sample collection, a network was constructed. Each artist formed a node and edges between these artists indicated the above collaboration. Collaboration simply indicated at that artist has registered another artist's involvement in a track, which is not always bidirectional; for example, The London Symphony orchestra has recent music `collaborating'  with Wolfgang Amadeus Mozart, which is neither reciprocated or indicative of a working relationship between the artists. It does however indicate a connection between these two artists in terms of a commonality of music.

These collaborations are collapsed into an undirected and unweighted graph with 3,766,631 edges.  The network formed from this graph and the associated metadata about artist names, genres and popularity, forms the largest connected component of the Spotify artist collaboration network and represents the majority of modern music artists. 

\subsection{Network Statistics}

\begin{figure}[h!]
	\centering
	\includegraphics[scale=0.8]{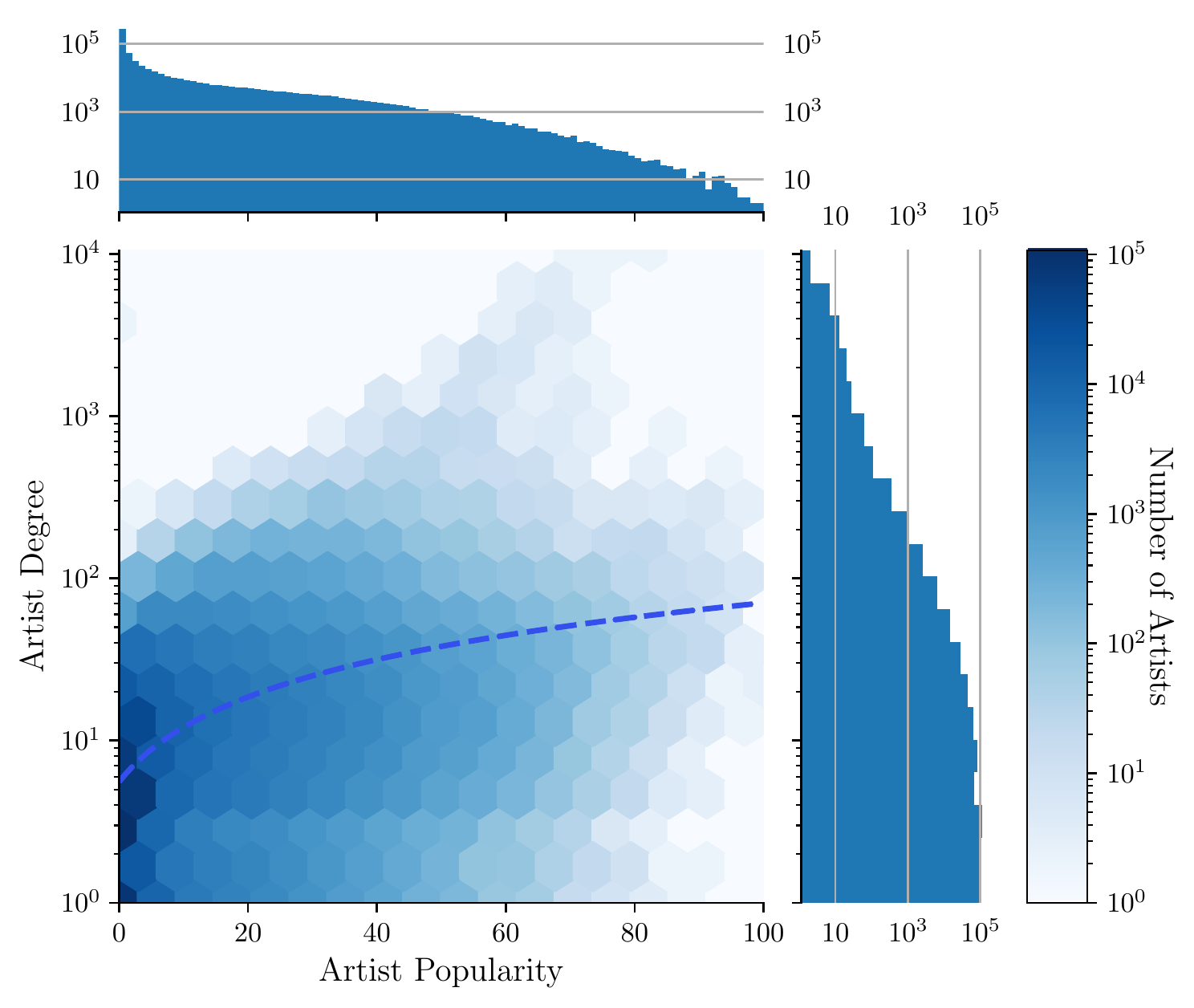}
	\caption{ Relationship between the popularity and degree of each artist in the Spotify artist collaboration graph. A linear trend line is fitted to the data shown as the blue dashed line. Artists with a higher popularity have more collaborations on average but high numbers of collaborations is not a determinant of popularity. The counts of artist collaborations have a long tail, with fewer than normal low-degree artists due to the snowball sampling. }
	\label{fig:PopDegreeScatter}
\end{figure}

The distribution of the artist collaborations (artist degree) and the popularity are shown in
\autoref{fig:PopDegreeScatter}. The most collaborated-with artist, Wolfgang Amadeus Mozart, has degree 10,514. A power-law (with exponent 2.988) fits the degree distribution poorly, but we observe around 4 orders of magnitude of variability and so assert that the distribution is heavy-tailed. The snowball sampling strategy naturally biases against low-degree nodes (e.g., no degree 0 nodes are collected), and so we expect that there are additional low-degree artists, for instance on small components unconnected to the main component. However, we will explicitly be examining how sub-sampling this graph affects network metrics below. 

Artist popularity is a Spotify internal metric, based on content consumption, from 0 to 100 with 100 being the most popular artist, Ed Sheeran. There is an abundance of low popularity artists, with 281,800 artists having popularity 0. The popularity distribution (shown in the marginal of \autoref{fig:PopDegreeScatter}) has a resemblance to an exponential distribution with rate 0.37, but is truncated at a maximum popularity of 100. The variance at higher popularities is likely caused by the sampling becoming more noticeable with fewer artists.  

Artists are more likely to collaborate with other artists that have popularities similar to their own, with popularity homophily~\cite{newman_mixing_2003} of 0.250. This does not extend to an artist's degree, with a degree assortativity of -0.016. Further, the degree of an artist and their popularity have a positive Pearson's correlation coefficient of 0.203, indicating that more popular artists have additional influence through greater collaboration. 

Of the artists with metadata, not all are labelled into genres, usually correlated with the artists popularity and hence Spotify's incentive for correct metadata. Artists can have a large number of genres assigned to them to incorporate the variety of music produced. Of the edges between artists with available genre metadata, 47.96\% of the edges are between artists with at least one overlapping genre.

\section{Critical transitions in music centrality}

Of interest in social graphs is a measure of \emph{importance}. A typical choice for determining importance in a network is a graph centrality measure~\cite{bonacich_power_1987}. While many measures exist, eigencentrality is chosen as it is fast to compute on the large graph and relatively robust to changes in the network~\cite{niu_robustness_2015,costenbader_stability_2003,borgatti_robustness_2006}. Eigencentrality uses the idea that artists that are important share that importance with nodes they are connected to, and solves the eigenvector problem $Av = \lambda v$, where A is the adjacency matrix. The adjacency matrix has only non-negative entries; hence the Perron-Frobenius theorem~\cite{cvetkovic_spectra_1980} gives that only the greatest eigenvalue vector provides the centrality measure~\cite{newman_mathematics_2008}.

\begin{figure}[h!]
	\centering
	\begin{subfigure}{0.6\textwidth}
		\centering
		\includegraphics[width=\textwidth]{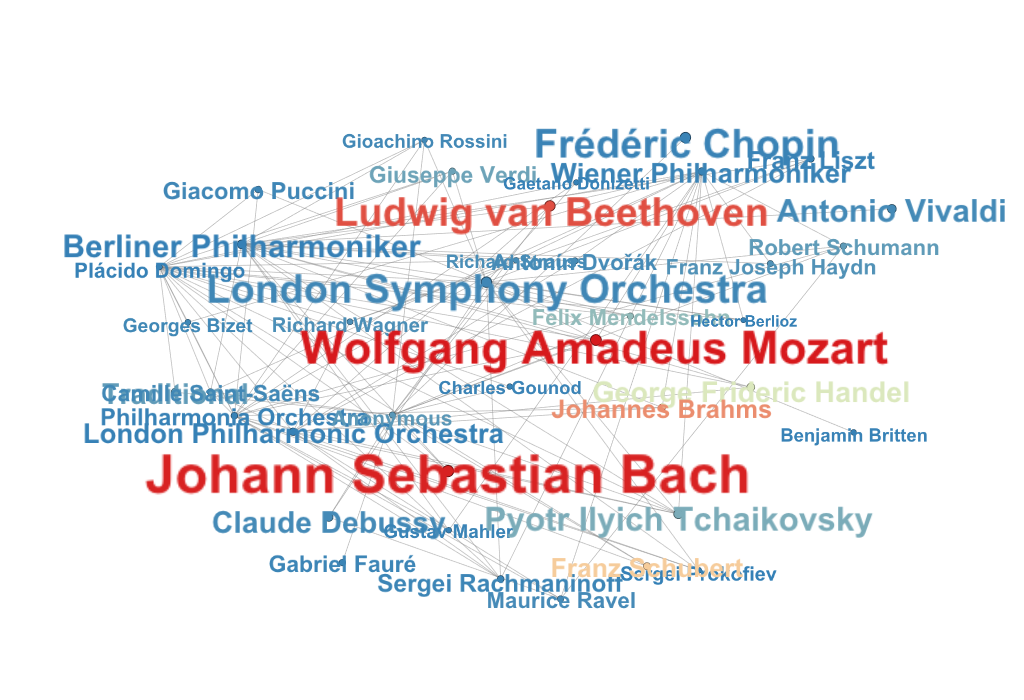}
		\caption{Full Network Centrality}
		\label{fig:fullnetwork}
	\end{subfigure}
	~
	\begin{subfigure}{0.6\textwidth}
		\centering
		\includegraphics[width=\textwidth]{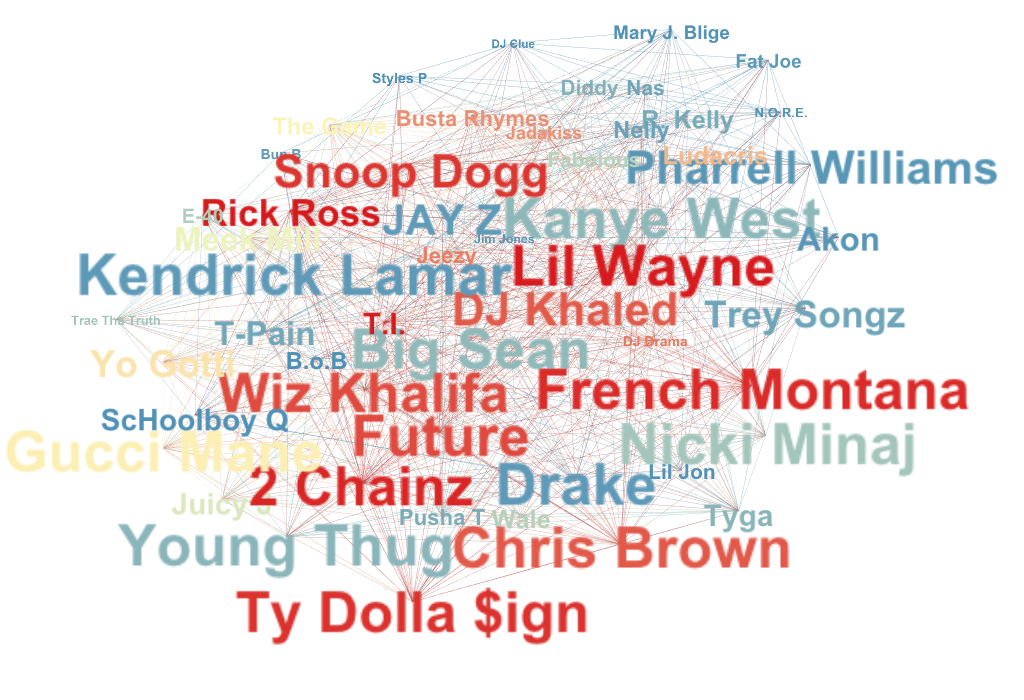}
		\caption{Most Popular Artists Centrality}
		\label{fig:popualirtynetwork}
	\end{subfigure}
	\caption{(a) Subgraph of network with highest eigencentrality in the full graph. Artists in this core are mostly classical artists. (b) Subgraph of network with highest eigencentrality when eigencentrality is calculated \emph{only} on the subgraph of high popularity nodes ($\geq$60). Artists in this core are characterised as rap artists. Colour is the relative centrality, with red being the most central; all artists shown have a higher relative centrality than most nodes. Size is proportional to popularity. The change in the most central core from classical artists to rap artist is observed after a critical transition in the graph at a popularity threshold of 46.}
	\label{fig:centernetworks}
\end{figure}
\begin{figure}[h!]
	\centering
	\begin{subfigure}{0.48\textwidth}
		\centering
		\includegraphics[width=\textwidth]{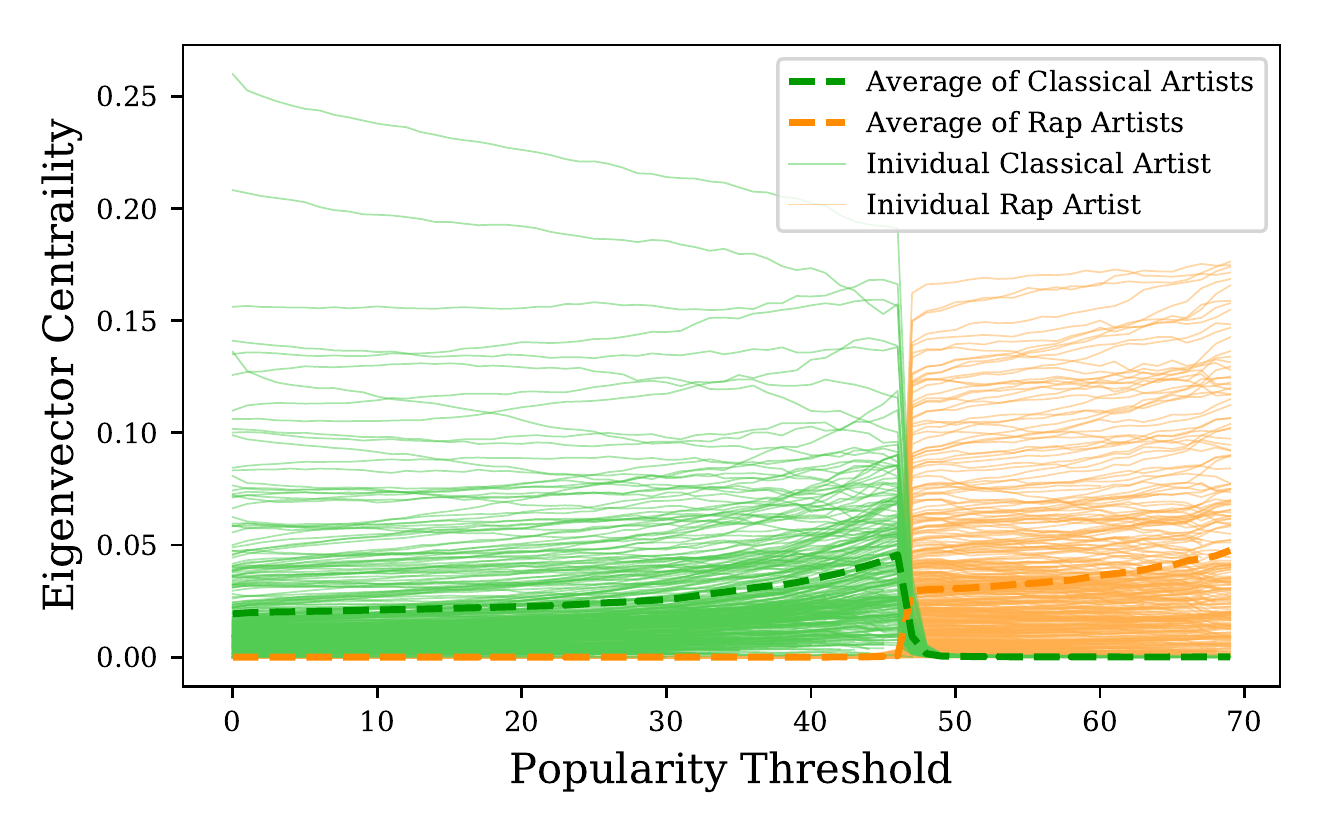}
		\caption{}
		\label{fig:criticaltransition}
	\end{subfigure}
	~
	\begin{subfigure}{0.48\textwidth}
		\centering
		\includegraphics[width=\textwidth]{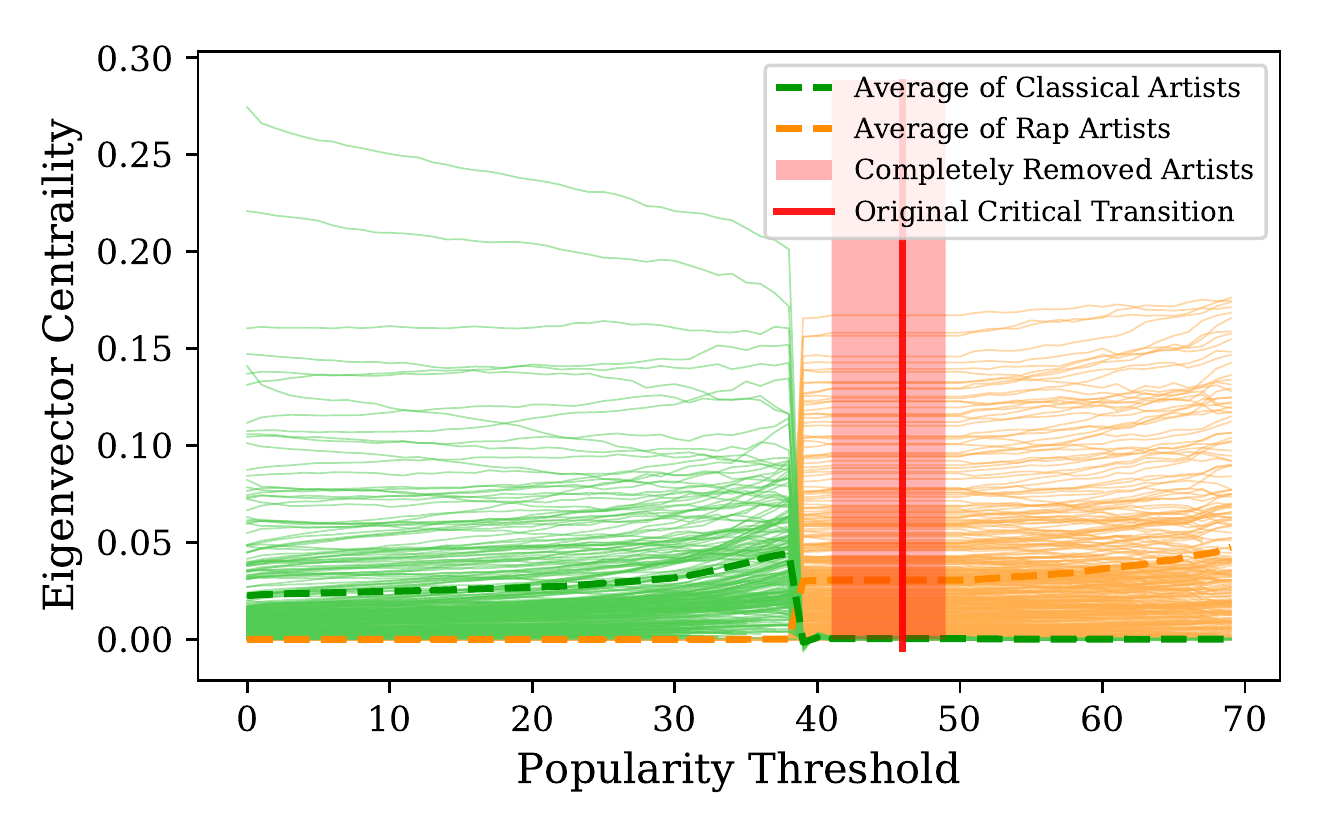}
		\caption{}
		\label{fig:real_removal}
	\end{subfigure}
	\caption{Change in centrality of all classical artists and all rap artists in the Spotify artist collaboration graph as popularity thresholding is applied. Artists of other genres have negligible centrality. In (a), the critical transition in centrality between the two groups can be seen at a threshold of 46. In (b), the artists with popularity in the red zone are removed from all threshold levels and the thresholding is repeated. A similar critical transition can be seen shifted left due to the removal.}
\end{figure}

We use eigencentrality to calculate the centrality of nodes over the full network. Artists with high centrality are shown as a subgraph in \autoref{fig:fullnetwork}, with Wolfgang Amadeus Mozart being the most central artist followed closely by Johann Sebastian Bach. These centrality results are similar to results of centrality metrics applied to the network of classical music composers~\cite{park_topology_2015}, although differ slightly due to the multi-genre nature of this data. The high number of connections of classical artists in addition to the diversity and variety of those connections leads them to form the most central, and through extension most important, core of the musical network.

However, in many situations an entire network cannot be collected. A common comprise given this constraint would be to focus on nodes with a high popularity. To focus on only popular artists we take the subgraph of artists with popularity greater than 60. The eigencentrality of this graph is found and artists with high centrality are shown similarly in \autoref{fig:popualirtynetwork}. 

Surprisingly, when only considering high-popularity artists, the most central core is entirely different. It comprises the popular rappers, with the most central being Rick Ross. It is important to note that most major classical artists\emph{are still included in the network}: Mozart, Bach, Tchaikovsky and many more have popularities well above 60, but the structure of the network has changed such that the centrality has shifted towards the rappers.

We measure the effect of taking subgraphs of only artists above the popularity threshold for a variety of values. In \autoref{tab:examplethresholds}, we highlight the most central artists at different threshold values. For low popularity thresholds up to 46, the most central artists are all classical artists. 

As the thresholding value increases there are small changes in the relative ranking of each artists centrality, such as the swap in rank of Bach and Beethoven. However, at a popularity threshold of 47 the artist centralities undergoes a critical transition, where the ranks of artists changes \emph{en-masse} (\autoref{fig:criticaltransition}).  In this critical transition, all classical artists lose almost all centrality and the rap artists rise to dominate the high centrality ranks. The rap artists then continue to have the highest centrality ranks for the remainder of all thresholding. 

This critical transition is not caused by the removal of any key classical artists or of a single artist otherwise, but rather by a broader change in network structure. To demonstrate this the thresholding approach is repeated on a new version of the graph, wherein all artists with popularity between 40 and 50 have been removed from the graph entirely. This new experiment shown in \autoref{fig:real_removal} demonstrates that even without the artists we once removed near the critical region, the critical transition still occurs.

The genre classifications provided here are selected to be representative. For example, although `2 Chainz' is classified by Spotify as belonging to a number of specific genres including `\emph{dwn trap}', `\emph{pop rap}', `\emph{southern hip hop}', and `\emph{trap music}'; we classify him more broadly as `\emph{rap}' to simplify.

\begin{table}
  \centering
  \begin{tabular}{l|lc|lc}
  \toprule
  \begin{tabular}[c]{@{}l@{}}%
  Popularity~\\Threshold\end{tabular} 
  & \multicolumn{2}{c|}{0}               & \multicolumn{2}{c}{60}                \\
  \hline
  Rank &            Name & Genre &            Name & Genre \\
  1 &  Wolfgang Amadeus Mozart &  classical &       Rick Ross &   rap \\
  2 &    Johann Sebastian Bach &  classical &            T.I. &   rap \\
  3 &     Ludwig van Beethoven &  classical &       Lil Wayne &   rap \\
  4 &           Franz Schubert &  classical &        2 Chainz &   rap \\
  5 &           Giuseppe Verdi &  classical &  French Montana &   rap \\
    \hline
  \end{tabular}
  
  \begin{tabular}{l|lc|lc}
    \toprule
    \begin{tabular}[c]{@{}l@{}} %
    Popularity~\\Threshold\end{tabular} 
    & \multicolumn{2}{c|}{46}               & \multicolumn{2}{c}{47}                \\
    \hline
    Rank &            Name & Genre &            Name & Genre \\
  1 &   Wolfgang Amadeus Mozart &  classical &       Rick Ross &   rap \\
  2 &      Ludwig van Beethoven &  classical &       Lil Wayne &   rap \\
  3 &     Johann Sebastian Bach &  classical &            T.I. &   rap \\
  4 &           Johannes Brahms &  classical &        2 Chainz &   rap \\
  5 &  Pyotr Ilyich Tchaikovsky &  classical &  French Montana &   rap \\
    \hline
  \end{tabular}
\caption{The highest centrality artists in the Spotify Collaboration Graph at different levels of popularity thresholding.  The artists with the highest eigenvector centrality undergo a critical transition between the popularity thresholds of 46 and 47, whereas the centrality ranking has minors changes in the ranges of 0 to 46 and 47 to 60. This contrasts the relative stability of the centrality during most threshold increases with the sudden change \emph{en masse} during the 46-47 popularity change, in which the artists that form the most central core of the network swap from classical artists to rap artists.}  \label{tab:examplethresholds}
\end{table}

\section{Social Group Centrality Model}

We propose a model of social group centrality (SGC) as a simplified way of capturing the dynamics of critical changes in centrality under thresholding. The SGC model consists of three groups; ``celebrities'', ``community leaders'' and ``the masses.'' The masses ($M$) are the largest group consisting of a randomly generated Barabási–Albert graph. They have random popularity assigned to each node according to an exponential distribution with rate 20 and no explicit homophily introduced, $\text{pop}(v_M) \sim \text{Exp}(20)$, where $v_M$ are vertices in the masses, $M$. The popularity is then truncated to a maximum of 100 for comparison purposes with the Spotify graph.

We then attach the remaining two groups. The celebrities ($v_{celeb}$) are added as a clique, with each celebrity being uniform randomly attached only to nodes of popularity \emph{greater than} some value $k$ at a rate of $p_{celeb}$.  Conversely, the community leaders ($v_{leader}$) are added as a clique and attached uniformly to nodes of popularity \emph{less than} $k$ at a rate of $p_{leader}$, where $p_{leader} > p_{celeb}$. Both the celebrities and community leaders are assigned a high popularity to avoid removal during thresholding, $\text{pop}(v_{celeb}) = \text{pop}(v_{celeb}) = 100$.

\begin{figure}[h!]
	\centering
	\begin{subfigure}{0.48\textwidth}
		\centering
		\includegraphics[width=\textwidth]{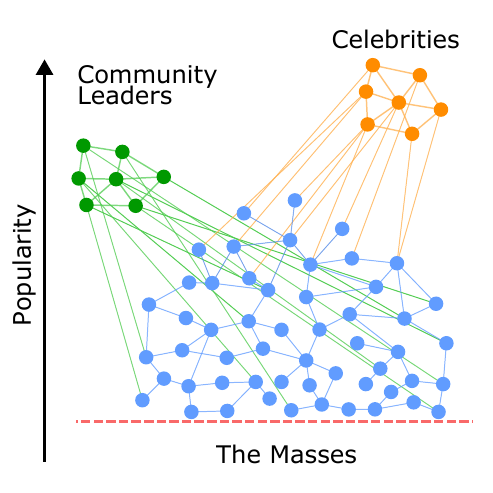}
		\caption{Full network}
		\label{fig:model_all}
	\end{subfigure}
	~
	\begin{subfigure}{0.48\textwidth}
		\centering
		\includegraphics[width=\textwidth]{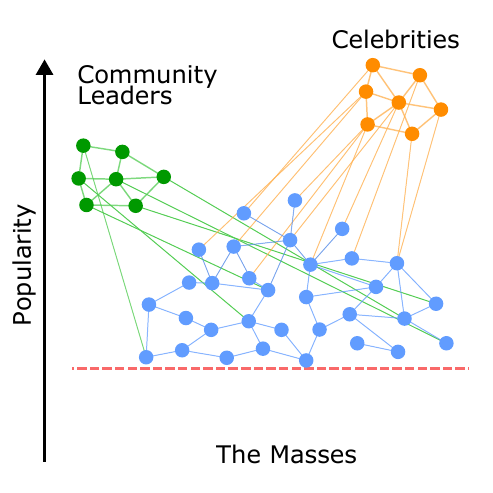}
		\caption{Popularity threshold applied}
		\label{fig:model_removed}
	\end{subfigure}
	\caption{A visual representation of the Social Group Centrality (SGC) model. Community leaders are randomly attached to lower popularity nodes in the masses, and celebrities are attached to fewer but higher popularity nodes. The red line shows a very weak popularity threshold. (b) shows how the network changes when a higher popularity threshold is applied removing more nodes. This network will have a change in centrality from the community leaders to the celebrities as an increasing threshold is applied.}
	\label{fig:model}
\end{figure}

This model is highly simplified, but demonstrates the dichotomy between grass-roots-style social influence - where impact is achieved through the connection of many low-influence individuals - and celebrity influence, where impact is not drawn directly from the cumulative network effects of many people, but rather the direct influence on the smaller group of most popular people. \autoref{fig:model_all} shows these connections visually, with the wide low lying connections of the community leaders and the concentrated connections of the celebrities to the high popularity masses.  By capturing these two types of influence patterns in the network model, we will demonstrate a method of disentangling them.

This model pertains to networks where there is both a connection network, and a measure of popularity. Often this comes in the pairing of social connections and consumption metrics. In Spotify this takes the form of the collaboration network and user-streaming based popularity, but could be applied to other social networks such as Twitter using friend networks (reciprocal followers) and follower counts (a measure of popularity of consumption). In both examples, it can prove difficult to collect the entire social graph, while relatively easy to collect the most popular nodes and their social connections.

\subsection{Changing Eigenvector Centrality}

\begin{figure}[h!]
	\centering
	\begin{subfigure}{0.48\textwidth}
		\centering
		\includegraphics[width=\textwidth]{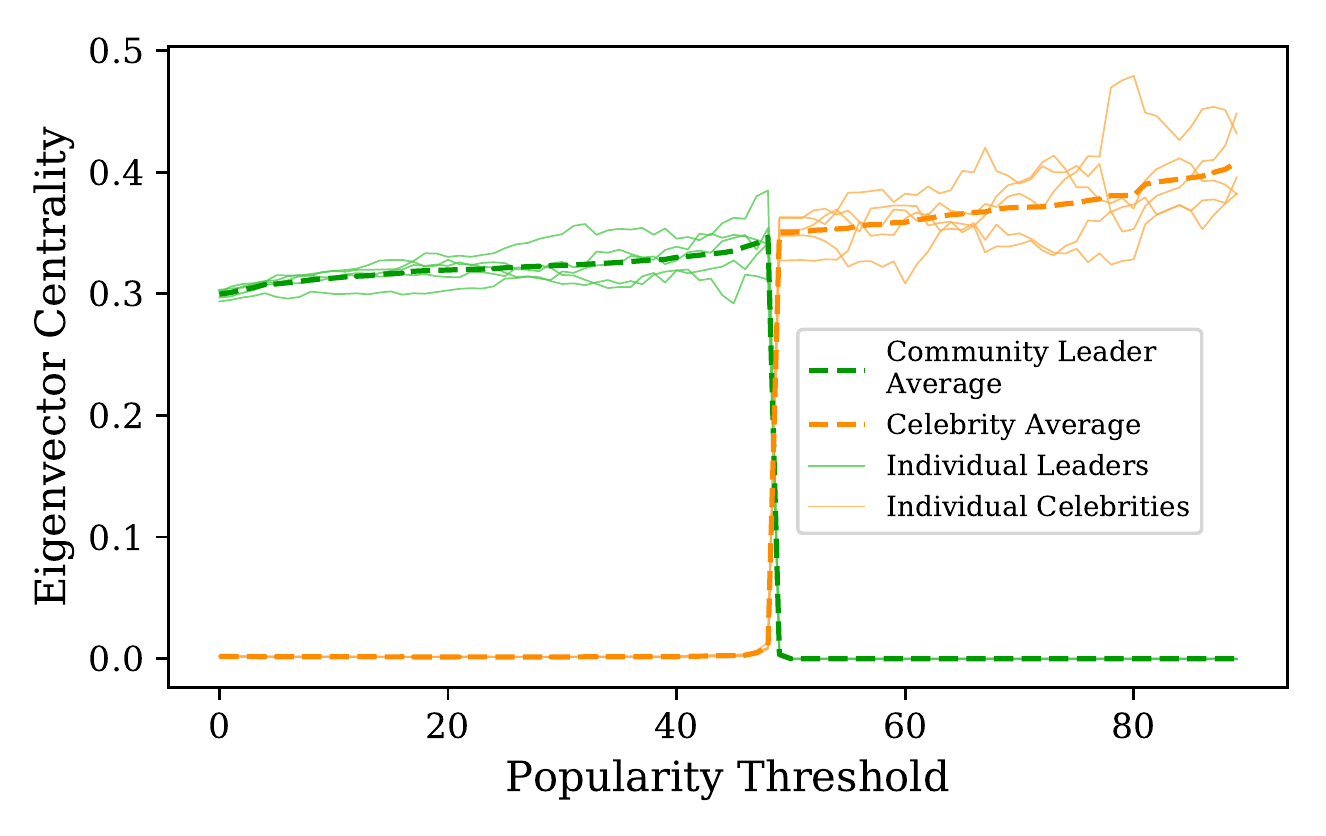}
		\caption{Centrality changes as low popularity nodes are removed from graph.}
		\label{fig:single_simulation}
	\end{subfigure}
	~
	\begin{subfigure}{0.48\textwidth}
		\centering
		\includegraphics[width=\textwidth]{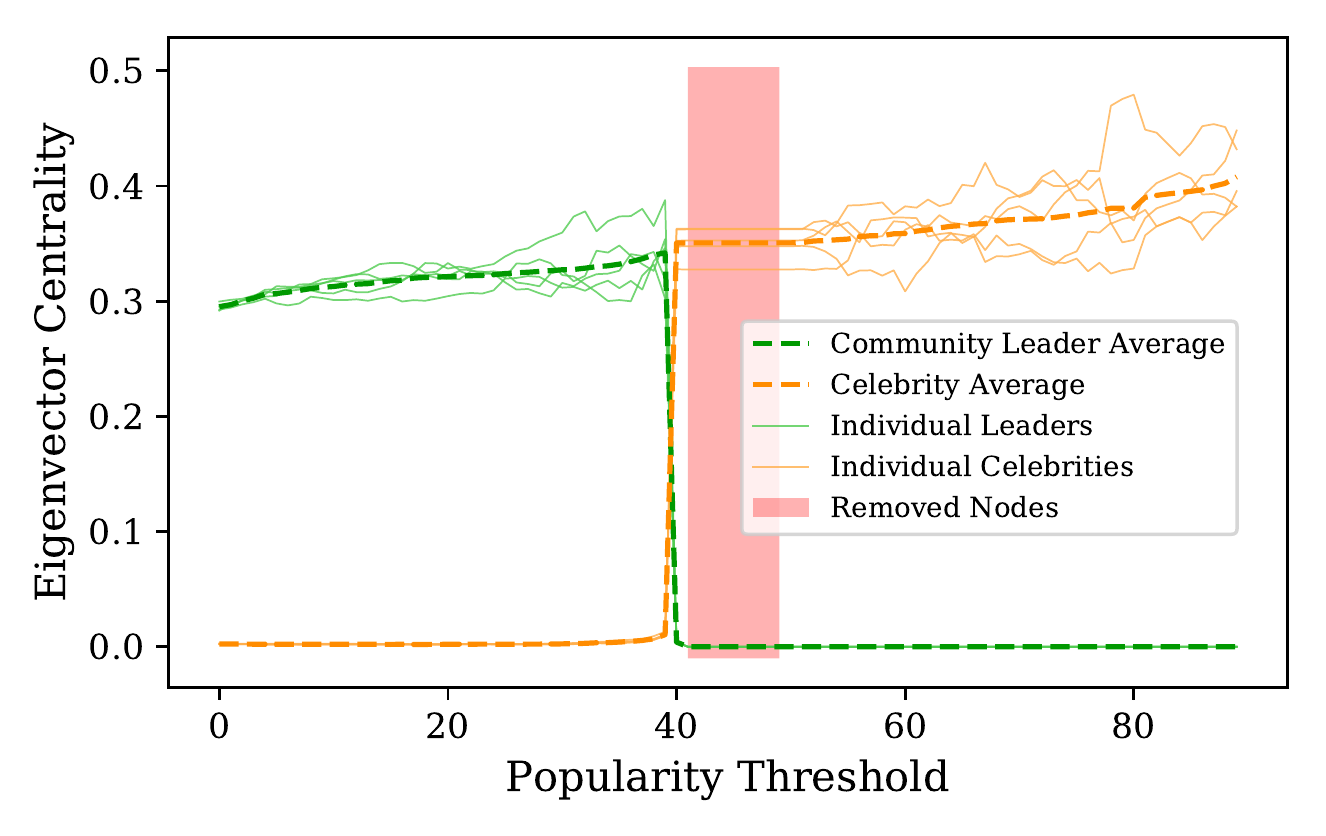}
		\caption{Parameters of transition or degree change analysis.}
		\label{fig:removed_simulation}
	\end{subfigure}
	\caption{Change in centrality of all community leaders and celebrities in a realisation of the Social Group Centrality model as popularity thresholding is applied. (a) shows a critical transition in centrality between the two group similar to the real data in \autoref{fig:criticaltransition}. In (b), the nodes with popularity in the red zone are removed from all threshold levels and the thresholding is repeated; identically to \autoref{fig:real_removal}. A critical transition still occurs, but shifted to the left.}
\end{figure}

Studying the SGC model using the same popularity-biased node removal gives similar critical transitions as seen in the Spotify data. In \autoref{fig:single_simulation}, a simulation is run to create a random graph from the model using $k=50$, $p_{leader} = 0.1$, $p_{celeb}=0.01$, and $|v_{M}| = 10000$.  We measure centrality for subgraphs taken with only nodes above the popularity threshold. We again see a sharp critical transition in centrality between community leaders and the celebrities, while nodes in the masses have a negligible centrality.

When a graph is not constrained by a high popularity threshold, the large number of connections from the community leaders to the masses makes them highly central, which is further reinforced by the clique. As a popularity threshold is applied, these community leaders have fewer connections as their low popularity neighbours are removed. Near the critical transition point, the average degree on the community leaders becomes less than the average degree of the celebrities, leading to a change in centrality (explored further below).

The change in centrality between the two groups is a consequence of the model construction, but the critical nature of the transition is interesting. Similar to the experiment in the Spotify graph, the nodes near the critical region are removed from the graph entirely, and the popularity threshold centrality is recalculated at all levels. In \autoref{fig:removed_simulation}, this critical transition now occurs at a new lower popularity point, again indicating that the nodes in the critical region were not important connective nodes in the graph that caused the critical transition. Thus the property arises from global structural changes in the network.

\subsection{Changing Eigenvalues}

\begin{figure}[h!]
 	\centering
 	\begin{subfigure}{0.48\textwidth}
 		\centering
 		\includegraphics[width=\textwidth]{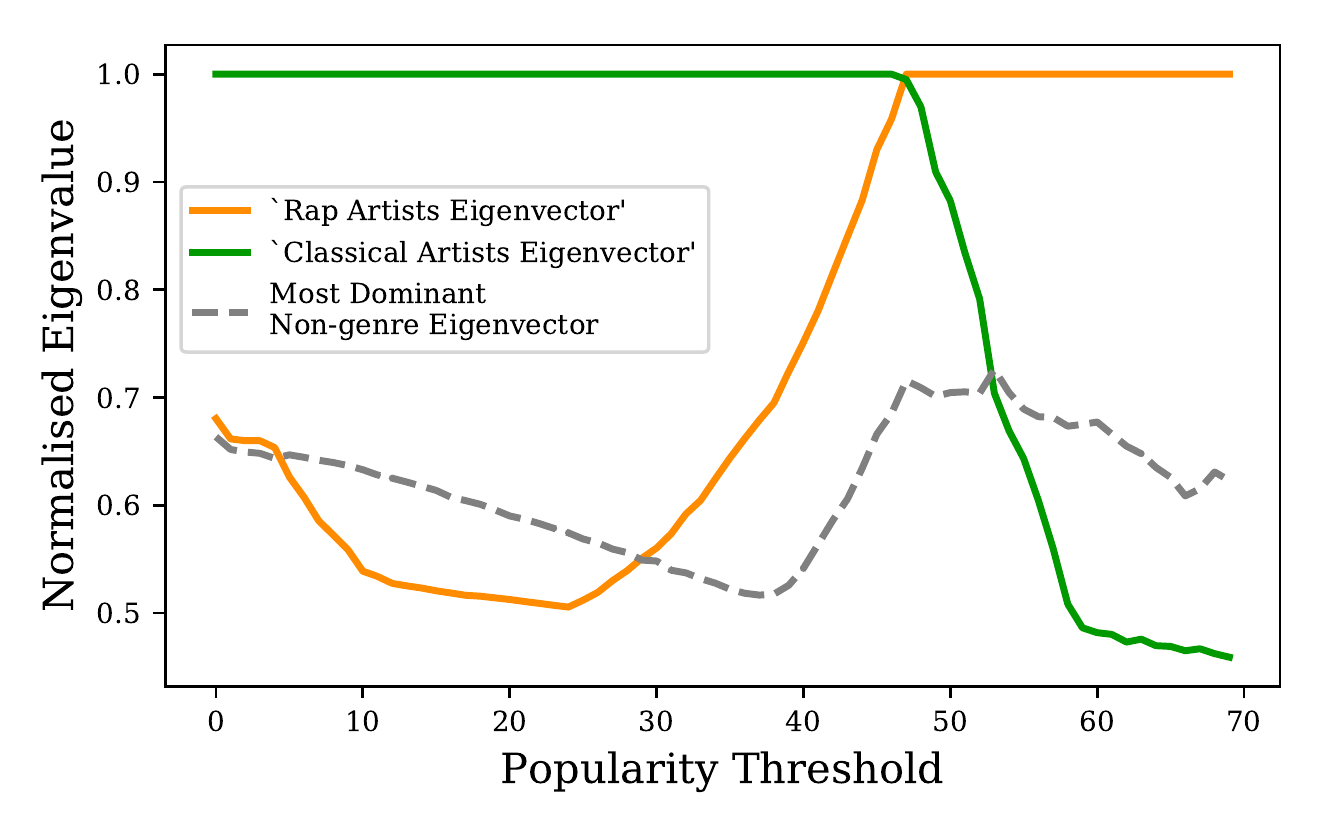}
 		\caption{Spotify Artist Collaboration Graph}
 		\label{fig:real_eigenvalues}
 	\end{subfigure}
 	\medskip
 	\begin{subfigure}{0.48\textwidth}
 		\centering
 		\includegraphics[width=\textwidth]{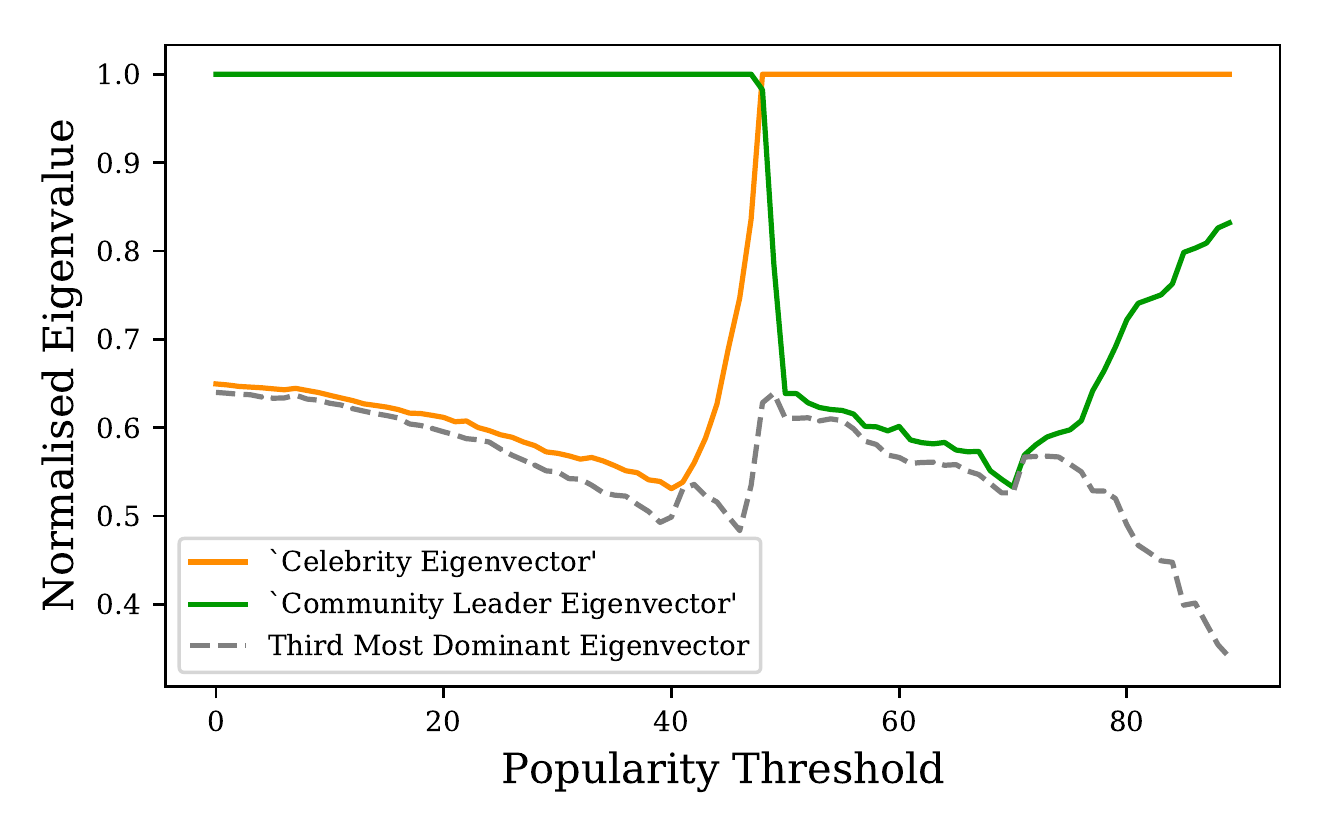}
 		\caption{Social Group Centrality Model}
 		\label{fig:sim_eigenvalues}
 	\end{subfigure}
 	\caption{Changes to most dominant eigenvalues of the adjacency matrix as popularity thresholding is applied to the graphs. Eigenvalues are normalised to the largest eigenvalue and are stylised according the group of nodes with high centrality in the corresponding eigenvectors. A swap between the dominant eigenvectors can be seen, corresponding to the critical transition in centrality.}
 	\label{fig:eigenvalues}
 \end{figure}

 The eigencentrality scores are drawn from the eigenvector with the largest eigenvalue (the dominant eigenvector). In both the Spotify artist collaboration graph and the SGC model, we see a change in the relative magnitudes of the largest eigenvalues at the critical transition. In \autoref{fig:eigenvalues}, as the threshold approaches the critical transition, the value of the second largest eigenvalue approaches the value of the largest eigenvalue. At the point of critical transition the eigenvectors swap, leading to the change in centrality. 
 
 \begin{figure}[h!]
 	\centering
 	\begin{subfigure}{0.48\textwidth}
 		\centering
 		\includegraphics[width=\textwidth]{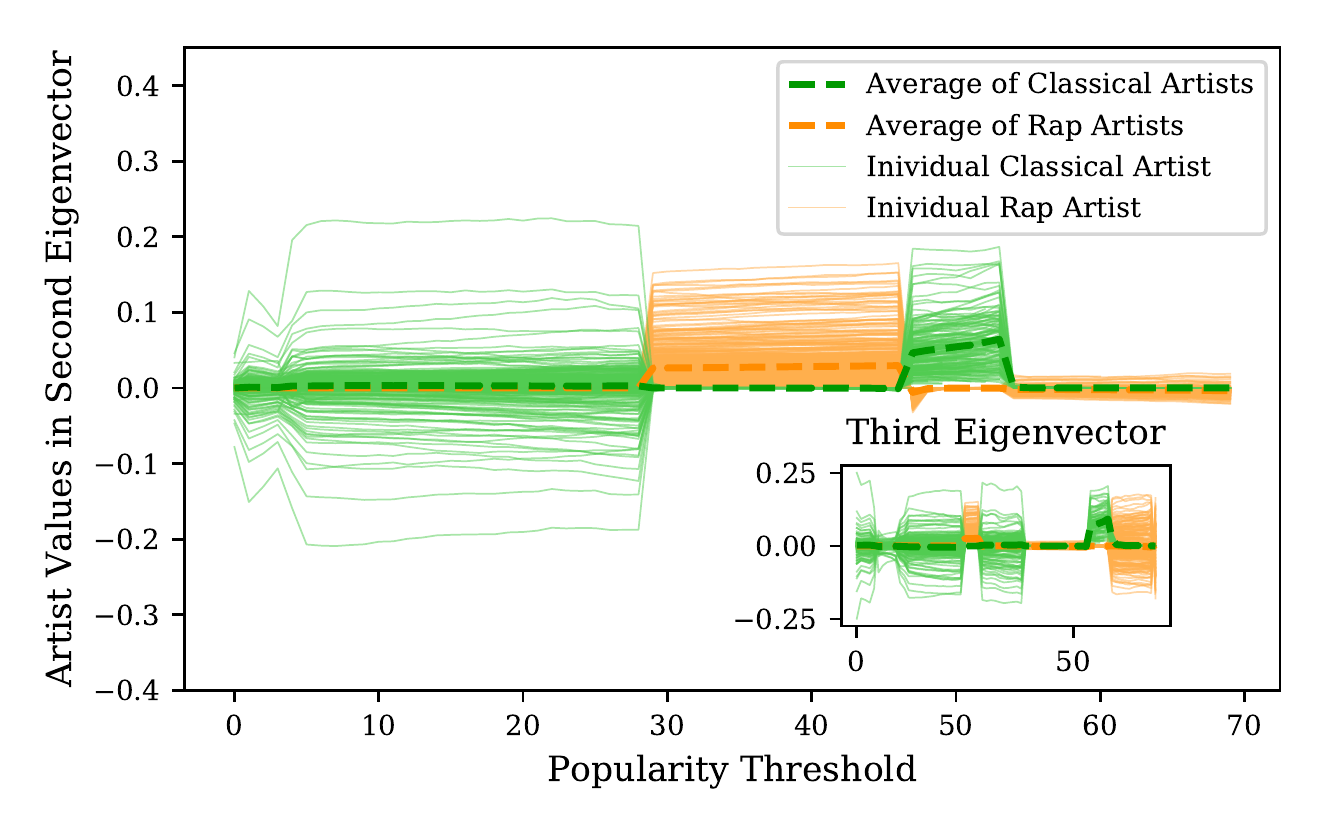}
 		\caption{Spotify Artist Collaboration Graph}
 		\label{fig:real_second}
 	\end{subfigure}
 	~
 	\begin{subfigure}{0.48\textwidth}
 		\centering
 		\includegraphics[width=\textwidth]{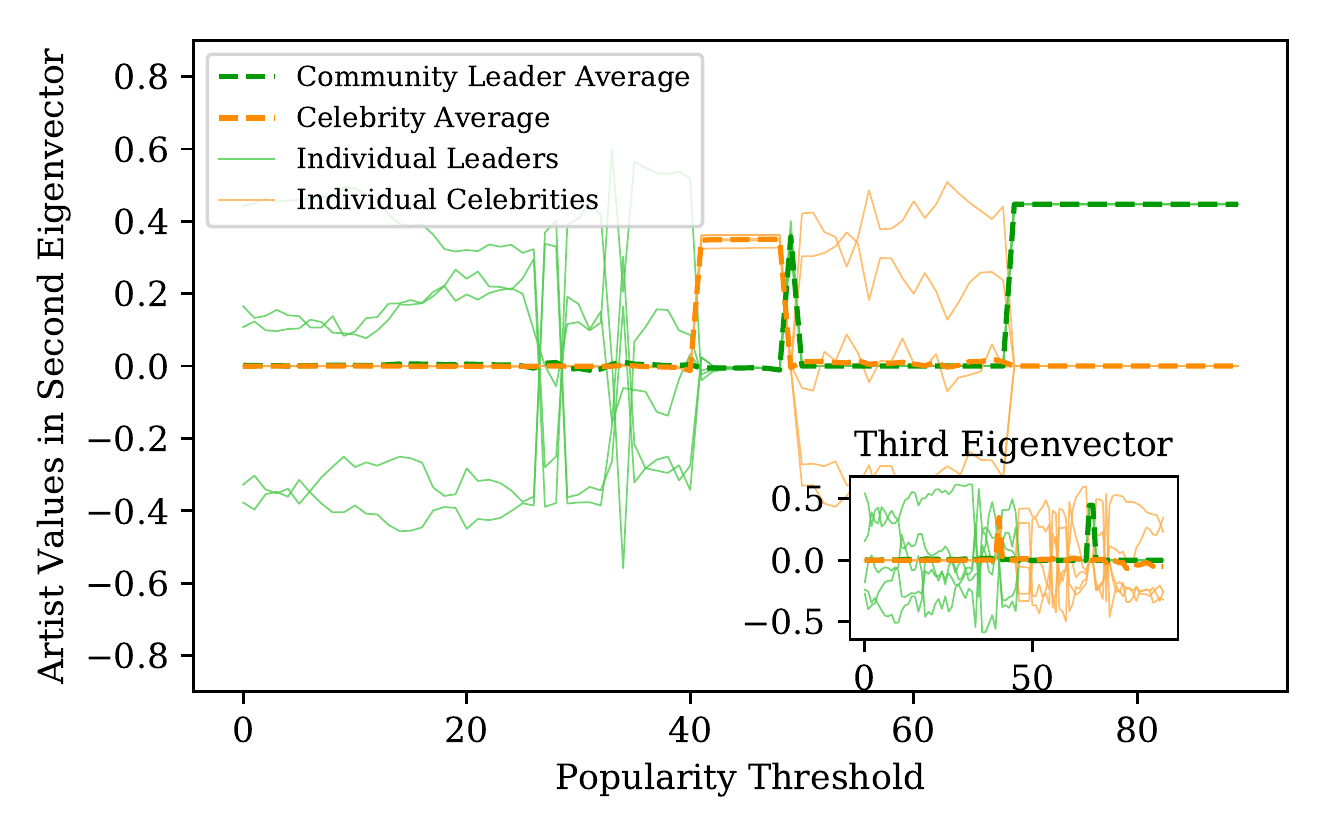}
 		\caption{Social Group Centrality Model}
 		\label{fig:sim_second}
 	\end{subfigure}
 	\caption{Change in second and third most dominant eigenvectors of the adjacency matrix as popularity thresholding is applied to the graph. Nodes are coloured according to their group in the graph. Swaps between the dominant and second eigenvectors can be seen at the critical transition points in the eigenvector centrality, with additional swaps between the second and third most dominant eigenvectors as eigenvalues change. This indicates that before the critical transition, the centrality of the ensuing node is observable in the second most dominant eigenvector.}
 	\label{fig:second_eigenvectors}
 \end{figure}
 
This swap in eigenvectors can also be seen in the lower-order eigenvectors. Looking at the elements of the second eigenvector corresponding to each artist, the critical transitions can be seen in \autoref{fig:second_eigenvectors}. Before becoming the most central group, the rap artists and celebrity are `incubating' in the second eigenvector until the point of critical transition. Additional critical transitions can also be observed between the second and third largest eigenvalues, which similarly correspond to critical changes in the eigenvectors. 

These second and third eigenvectors are not constrained by Perron-Frobenius to be non-negative, and indeed the second and third eigenvectors both contain non-negative values. However, near the point of critical transition, the second most dominant eigenvector resembles the reverse of the centrality result of the dominant eigenvector, in both the SGC model and the real data.

This phenomenon has been observed before in the context of the HITS algorithm for centrality, where a small gap between the eigenvalue of the dominant eigenvalue and the second eigenvalue can lead to instability in the centrality metric from a similar swapping of eigenvectors~\cite{ng_link_2001,ng_stable_2001}. 

\section{Causes of the critical transition}\label{sec:causes}

In the SGC model, the threshold at which centrality swaps can largely be explained by the change in node degrees as the graph undergoes popularity thresholding. While this does not explain the critical nature of the transition, it does allow for control over \emph{where} the transition occurs. 
\begin{figure}[h!]
	\centering
	\begin{subfigure}{0.43\textwidth}
		\centering
		\includegraphics[width=\textwidth]{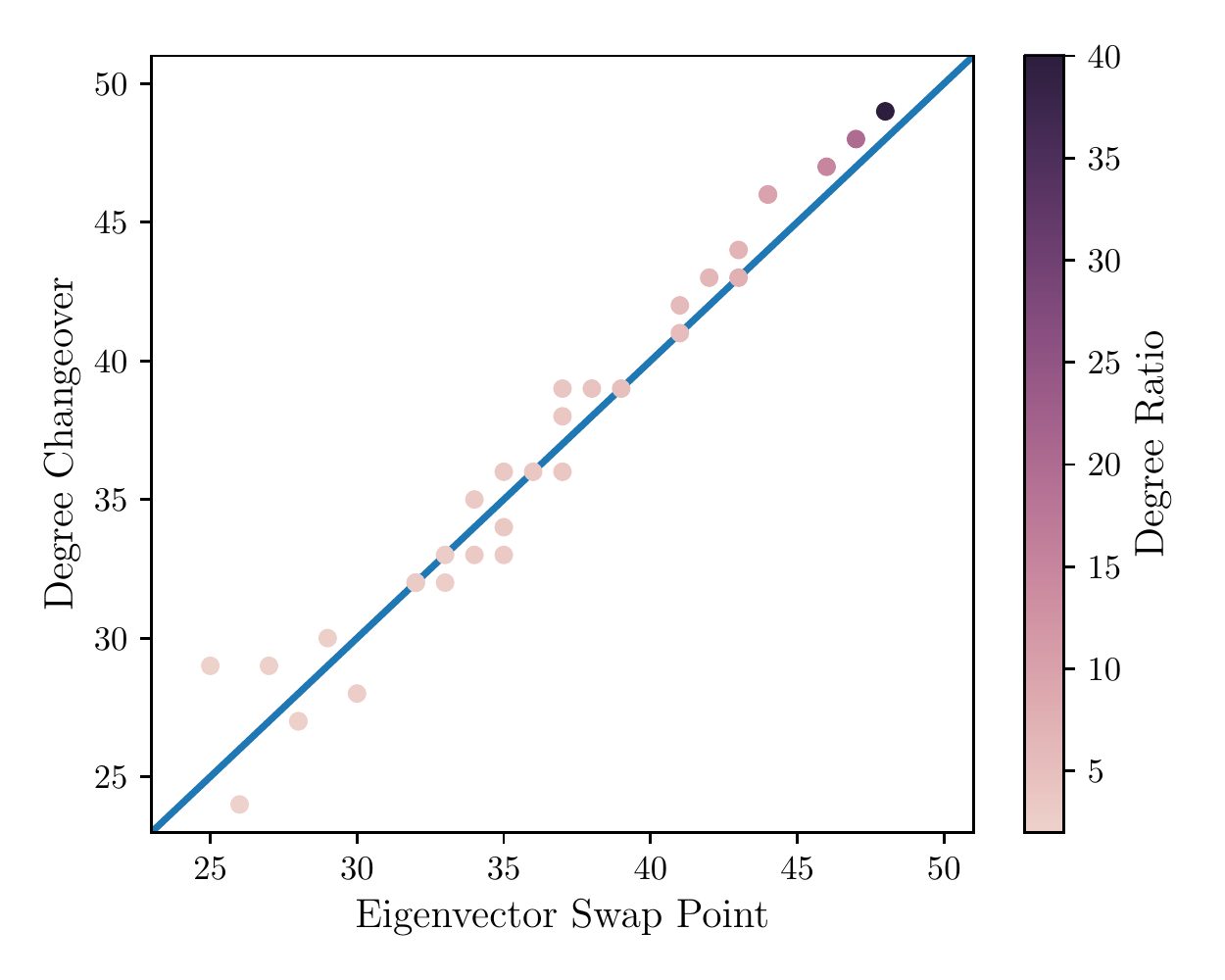}
		\caption{}
		\label{fig:degree_change_swap}
	\end{subfigure}%
	~
	\begin{subfigure}{0.55\textwidth}
		\centering
		\includegraphics[width=\textwidth]{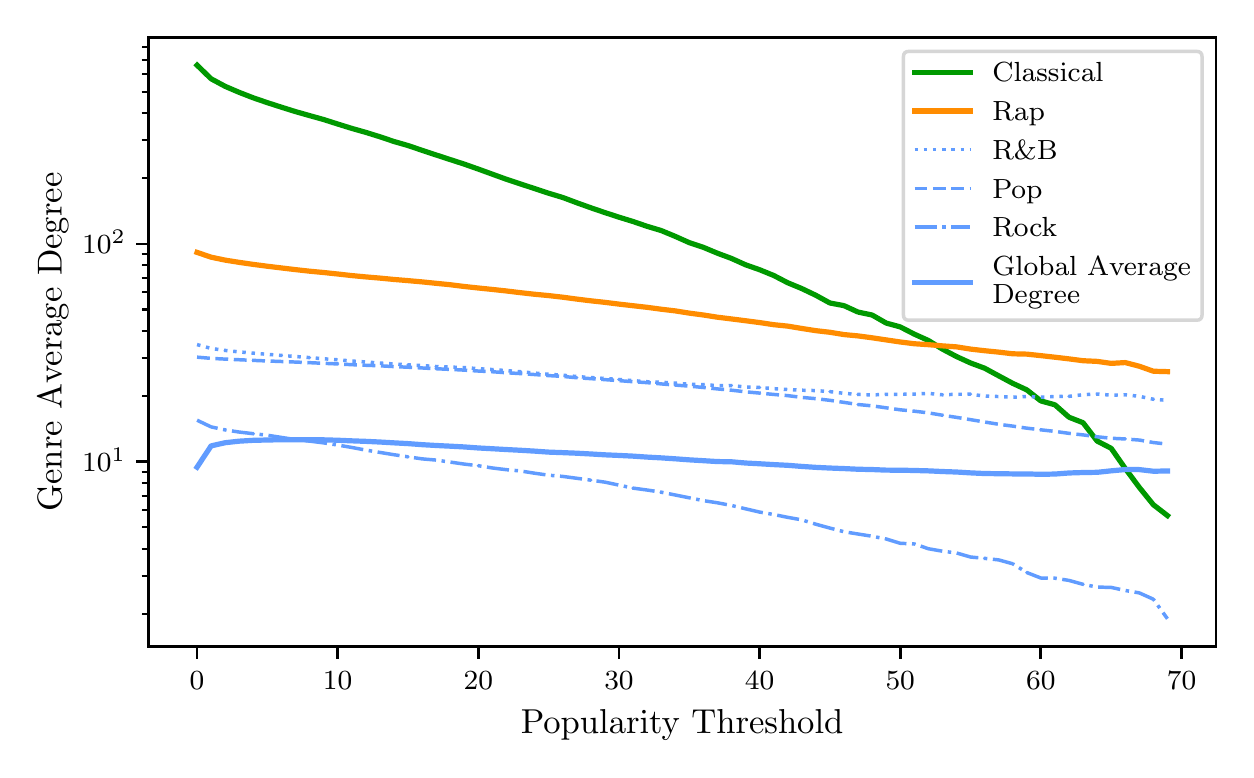}
		\caption{}
		\label{fig:real_degree_change}
	\end{subfigure}
	\caption{ (a) Relationship between the point at which celebrities first have higher average degrees than community leaders and the transition point in centrality for SGC model realisations with various starting ratios of group degrees. (b) The average degree of artists belonging to different genres in the Spotify collaboration graph. A strong relationship between the degree changeover and the critical transition is observed in the SGC model but these two events do not occur simultaneously in the Spotify collaboration network.}
	\label{fig:degree_change}
\end{figure} 

We explain the change in centrality using a mean value argument. The group of nodes corresponding to the masses, $M$, is generated as a Barabási-Albert random graph. We take that the centralities of these nodes are distributed according to an unknown distribution, $v^{eig}_{M}$. Recalling the relative definition of eigenvector centrality in \autoref{eq:eigenvector}, the centrality of an isolated community leader or celebrity would then naively be a sum of the centralities of the nodes it is attached to. We can hence expect the centrality of our leaders and celebrities to have a naive initial centrality of $d_{leader} E[v^{eig}_{M}]$ and $d_{celeb} E[v^{eig}_{M}]$, where $d$ is the degree of a node.

This naive centrality would then reinforced through the clique structure of the community leaders and celebrities, reaching an equilibrium with higher centrality than if the nodes were isolated. Importantly, this reinforcement is driven by the structure of the group, which is shared by both the community leaders and celebrities. As such, the expected centralities of the groups are proportional to their average degrees,  $E[v_{leader}^{eig}] \propto \bar{d}_{leader} $ \& $E[v_{celeb}^{eig}] \propto \bar{d}_{celeb} $.
 
While random structure in the graph will alter this result, we expect the point at which the celebrities surpass the community leaders in centrality to be tightly correlated with the point of average degree change. Indeed, an experiment run is shown in \autoref{fig:degree_change_swap}, where the ratio $\bar{d}_{leader}  / \bar{d}_{celeb} $ in the model is altered, demonstrating the correlation between the point of degree changeover and centrality change. High degree ratios result in the degree changeovers occurring very near the popularity range of 50. This leads to less variance between the degree changeover and eigenvector swap point compared to lower ratios; as lower ratios have less edges removed at each threshold, exposing the model to more random changes in centrality due to structure.

This simple model doesn't fully extend to the larger and more complex real data. In the Spotify artist collaboration graph the classical and rap genres have the highest average degree, shown in comparison to other large genres in \autoref{fig:real_degree_change}. The classical and rap artists do have a swap in their rank of average degree, but this is further away from the critical transition point of centrality compared to the model results.

The Spotify artist collaboration graph still shares many common elements with the social group model. Classical genre artists initially have an average degree of 728.75 but an average neighbour popularity of 19.93, compared to an average degree of 93.93 and average neighbour popularity of 41.52 for rap artists.
The subgraphs of both the classical and rap genres are more density connected than average (0.024 and 0.020 compared to an average genre density of 0.008), although this is far from being a clique.

A further limitation is of this degree analysis approach is the confounding factor of release counts. Classical artists have on average existed for longer than modern artists, allowing more time for musical adaptations and posthumous collaborations or remixes. The number of total releases an artist features on is hence an important factor which can make it difficult to extract meaningful conclusions from the relationship between popularity and collaboration. To examine this, the number of albums or releases an artist appears on was also collected for a large subset of the artists ($n=456,245$). Indeed, the log of the number of total releases by an artist is strongly correlated with both the log of the number of neighbours ($r = 0.8372, p < 10^{-16}$) and the popularity ($r = 0.6233, p < 10^{-16}$). In part it is possible that more popular artists are better resourced and more incentivized to keep producing new albums and re-releasing old ones, such as through duplicate songs across singles, EPs, LPs, or remixes albums. This makes it difficult to disentangle the relationships between popularity and the degree of collaboration within this network. 

These release trends can also be seen in the genre grouping. The average total number of releases is greater for classical artists (mean $=70.98$, median $=16$) and rap artists (mean $=61.11$, median $=31$) than the typical artist (mean $=12.8$, median $=2$). However, in the case of classical artists the majority of the edges are not true collaborative relationships but rather connections established through the renditions of their work by orchestras, modern artists and other groups. This weakens the comparison of classical artists to community leaders, but also helps inform the importance of the timespan since an artists prominence on influence in the network.

\section{SGC model parameter exploration}\label{sec:components}

Inducing a critical transition in the graph relies upon the different popularity distributions of connections coming from the community leaders and celebrities. In the original SGC model above, a simple uniform $U(0,50)$  distribution is used for community leaders and a $U(50,100)$ is used for celebrities. To validate the model using different parameters, two experiments are run.

\begin{figure}[h!]
	\centering
	\includegraphics[width=\textwidth]{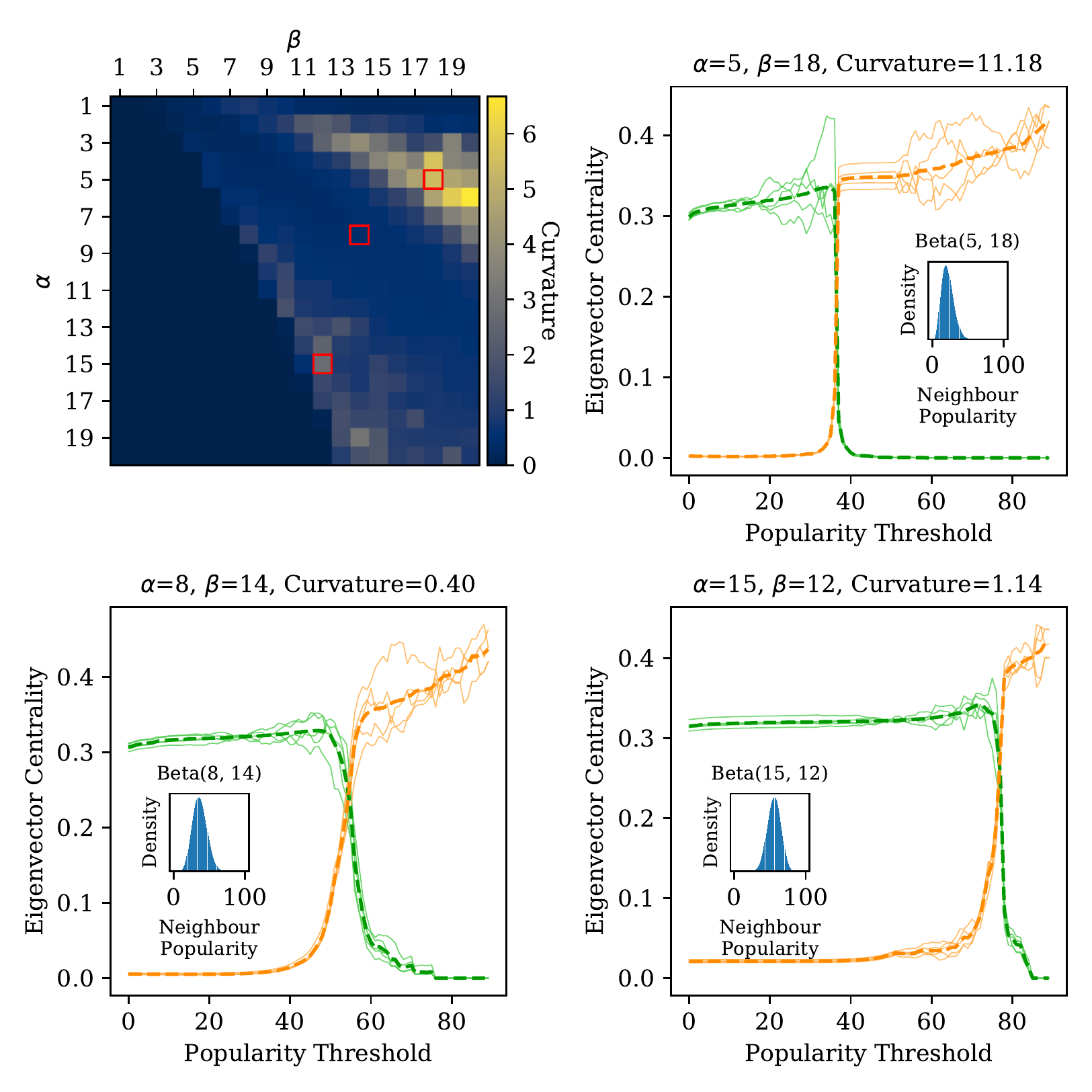}
	\caption{A Social Group Centrality model is modified such that community leader edges are connected to nodes of popularity chosen from a $Beta(\alpha, \beta)$ distribution. These models are simulated for various choices of  $\alpha$ \& $\beta$ and logistic curves are fitted to the average centralities of the two groups. The steepness parameter of the fitted curves are averaged across ten simulations at each set of parameters. The impact of $\alpha$ \& $\beta$ on curvature is shown for all parameter pairs with three example centrality changes highlighted with red squares and shown as individual plots. While any choice of parameters with $\alpha \leq 1.5\beta$ can induce a change in centrality, only a small region where $\beta$ is much greater the $\alpha$ produces the critical transitions of interest to this work. }
	\label{fig:beta_dist}
\end{figure}

In \autoref{fig:beta_dist}, the neighbour popularity distribution of $U(0,50)$ is replaced with a Beta distribution, $Beta(\alpha, \beta)$. For each pair of $\alpha$ \& $\beta$ parameters, ten simulations of the model are run. In each simulation a logistic curve is fitted to the average group centrality in the popularity threshold centrality curves. The absolute value of the logistic growth rate of the two fitted curves are averaged to produce a measure of \emph{curvature}. This curvature is related to the sharpness of the transition and is averaged over each of the simulations for each parameter pair. If no centrality transitions occur then the curvature is said to be 0. 

As the distribution changes for various parameter choice a few clear patterns emerge:
\begin{itemize}
  \item if $\alpha \geq 1.5\beta$, then the popularity distribution of community leaders is too left skewed, and they remain most central throughout all thresholding;
  \item as $\alpha$ and $\beta$ become close, critical transitions sometimes occur, shifted towards the higher end of popularity thresholding. At this end, there are less nodes remaining in the graph and critical transitions can be caused by key node removals;
  \item when $\alpha$ is slightly less than $\beta$, the change in centrality is smoothed, and edges of community leaders are removed at a similar time to the removal of celebrity edges;
  \item when $\alpha$ is much less than $\beta$, the distribution is right skewed -- similar to our $U(0,50)$ above -- resulting is critical transitions where the steepness of the fitted logistic is very high due to the almost vertical nature of the curve. 
\end{itemize}

\section{Other centrality measures}

For completeness of comparison, other centrality measures were explored applied to the Spotify artists collaboration graph and the SGC model. As previously stated the size of the graph means that several centrality measures such as betweenness and closeness are computationally intractable. However two relevant and related centralities are computable, PageRank and Degree centrality.

\begin{figure}[h!]
	\centering
	\begin{subfigure}{0.48\textwidth}
		\centering
		\includegraphics[width=\textwidth]{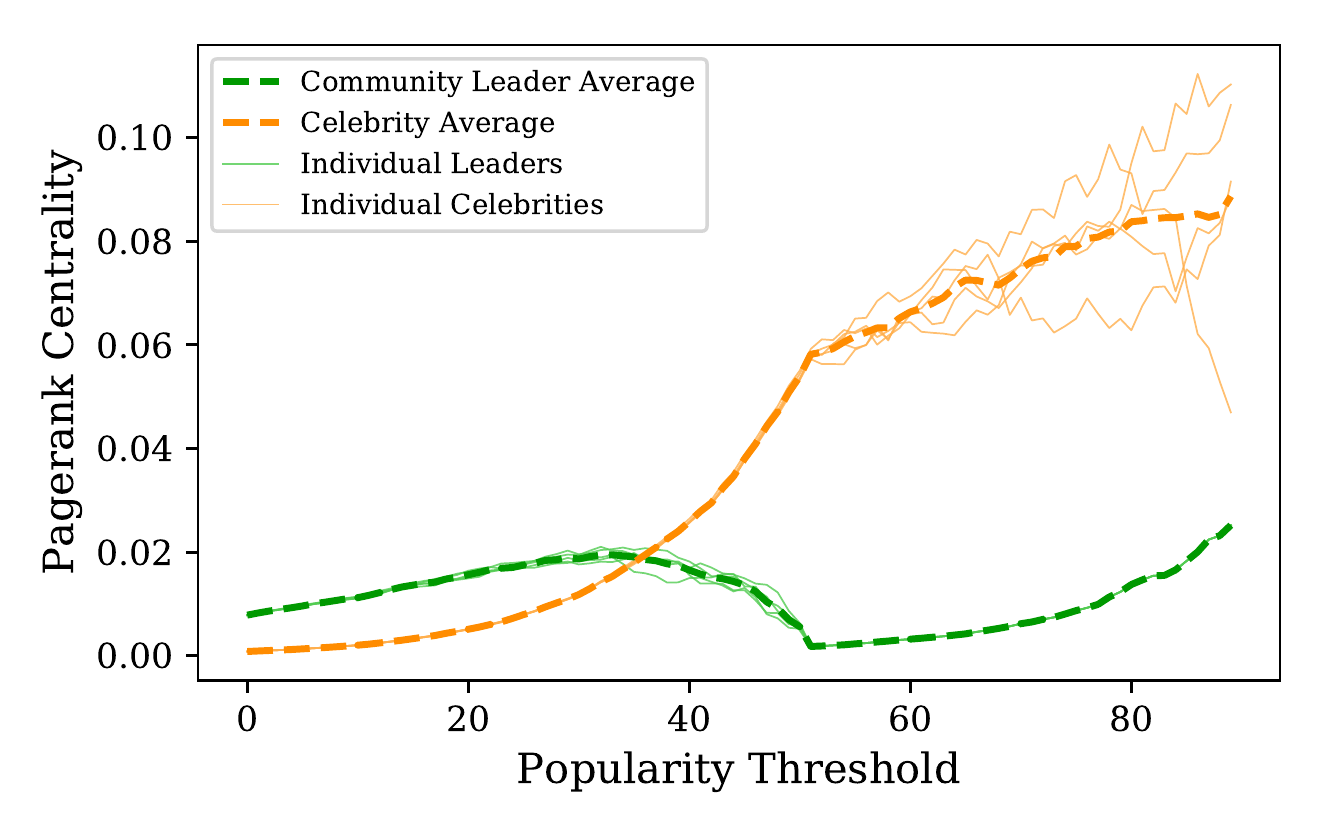}
		\caption{A simulation of SGC model}
	\end{subfigure}
	~
	\begin{subfigure}{0.48\textwidth}
		\centering
		\includegraphics[width=\textwidth]{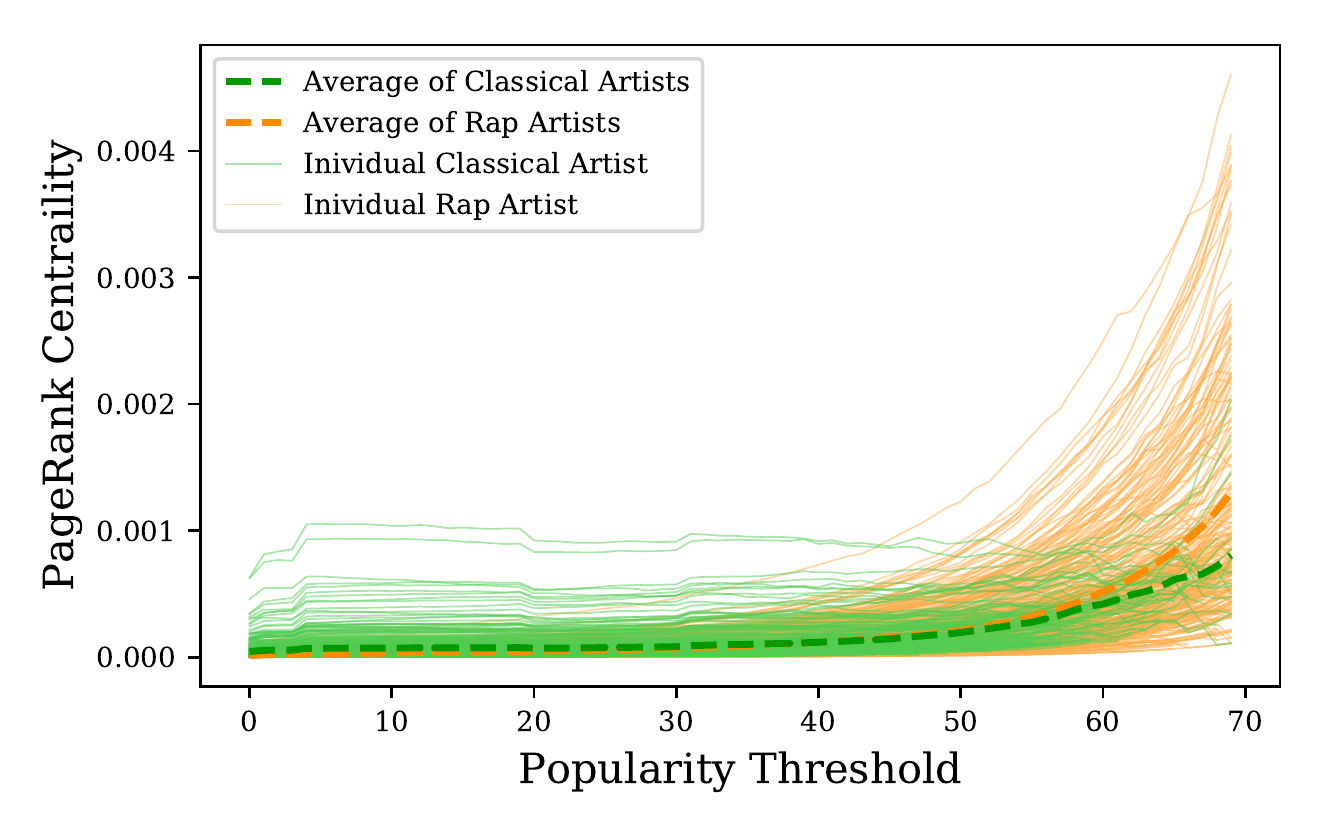}
		\caption{Spotify artist collaboration graph}
	\end{subfigure}
	\caption{Pagerank centrality under popularity thresholding for both the Social Group Centrality model and the Spotify artist collaboration graph. The Pagerank centrality still obverses a swap in the relative centrality of the groups in both contexts, but does so in a more gradual manner without a critical transition in both cases.}
	\label{fig:pagerank_centraility}
\end{figure}

In \autoref{fig:pagerank_centraility} we see that PageRank has an effect of smoothing the transition significantly. The end conclusions of the centrality are still similar -- that community leaders / classical artists start as a central core but celebrities / rappers become the central core in a popularity thresholded graph -- but the transitions between these conclusions is much smoother. The dampening effect of PageRank provides a `reset-to-uniform-distribution'~\cite{ng_link_2001}, which helps smooth this critical transition as has been observed before~\cite{ng_stable_2001}.

\begin{figure}[h!]
	\centering
	\begin{subfigure}{0.48\textwidth}
		\centering
		\includegraphics[width=\textwidth]{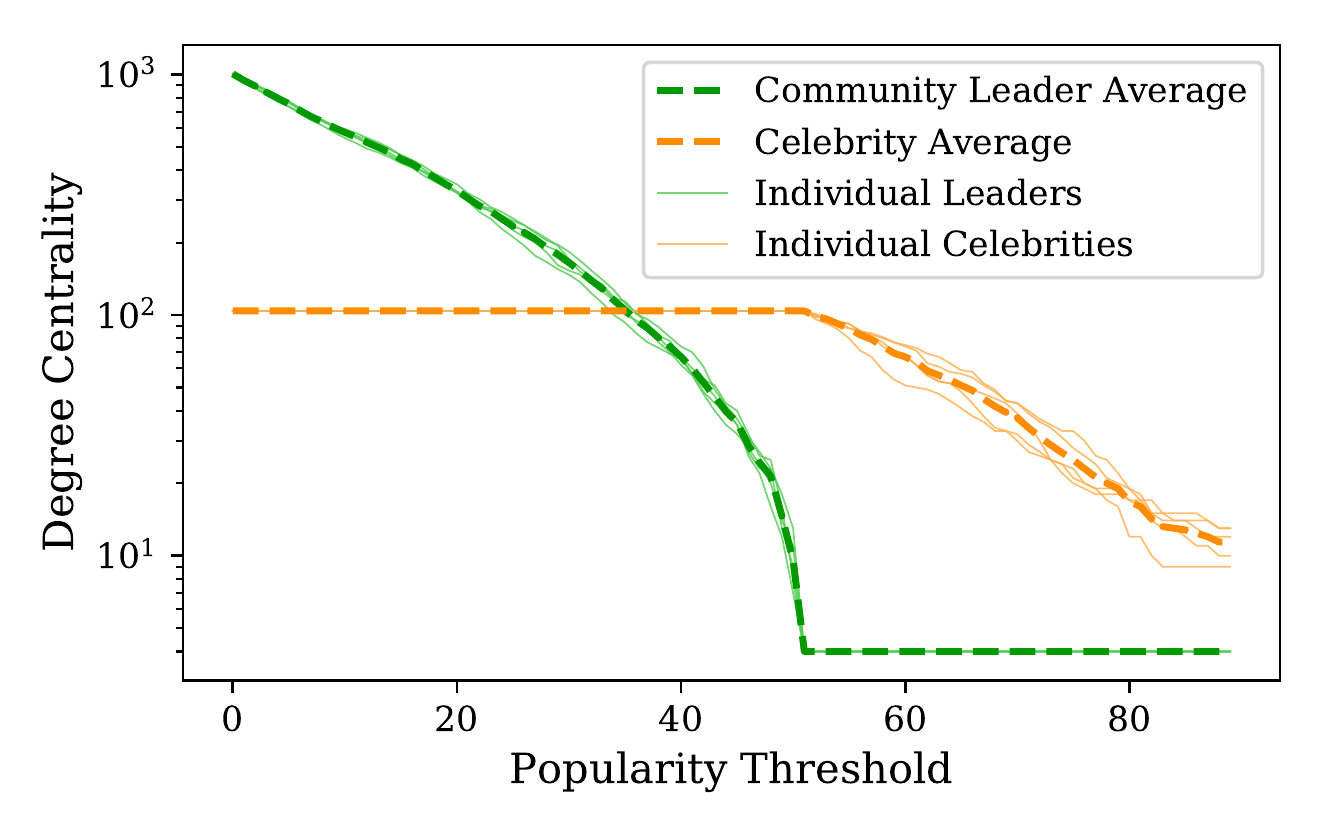}
		\caption{A simulation of SGC model}
	\end{subfigure}
	~
	\begin{subfigure}{0.48\textwidth}
		\centering
		\includegraphics[width=\textwidth]{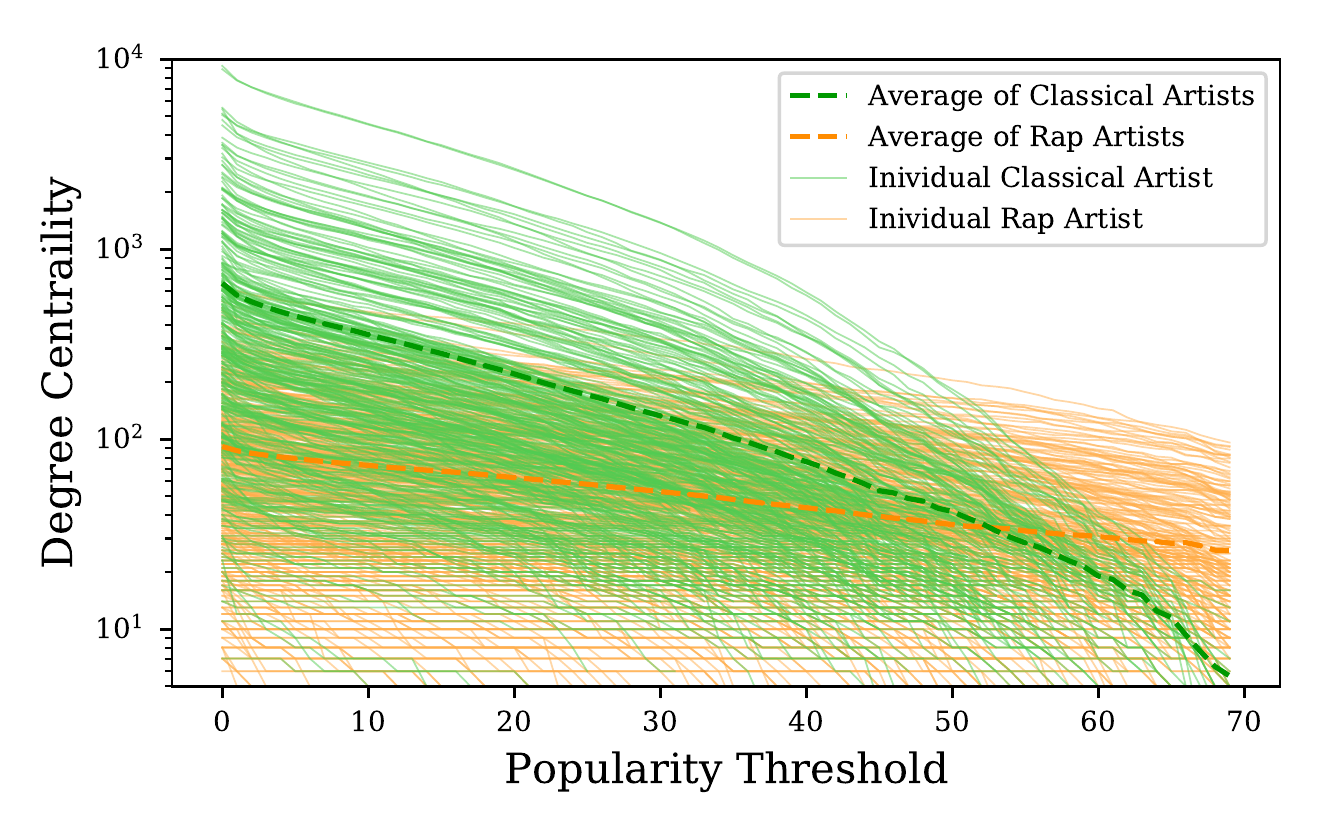}
		\caption{Spotify artist collaboration graph}
	\end{subfigure}
	\caption{Degree centrality under popularity thresholding for both the Social Group Centrality model and the Spotify artist collaboration graph. In both cases the two groups swap in degree centrality as thresholding is applied, but do so gradually and without a critical transition. This is in contrast to the critical transition of eigenvector centrality, which is influenced strongly by the degree of these nodes.}
	\label{fig:degree_centraility}
\end{figure}

Eigenvector centrality is itself an extension of degree centrality that takes into account neighbour centrality scores~\cite{newman_networks_2018}. In that sense, degree centrality shows us the causes of centrality change, without any interesting structural influence. In \autoref{fig:degree_centraility}, the natural changes in degree as low popularity nodes are removed is shown.

\section{Conclusion}

This work presents a new quantitative analysis of the music world, using data and metadata from the Spotify API to create a network of artists and their collaborations. The measure of an artists importance in the networks is probed using eigencentrality, resulting in two superficially antithetical inferences. We resolve these differences both through a descriptive analysis of the network structure and through the exploration of the critical transition and less dominant eigenvectors. 

From a descriptive perspective, the classical artists are the most central in the full artist network; with a large number of diverse connections to different genres and popularity levels. Conversely, the rap artists form the most central core of the  popular-artist only network, due to a high connectivity and bias towards collaborations with other high popularity artists.

The critical transition in the centrality, a phenomenon rarely seen, is caused by a swapping of the dominant eigenvectors. This transition is generated by a changing network structure rather than a singular major alternation in the network. The spectrum of the graph shows that as the critical transition approaches, the second largest eigenvalue approaches the first, warning of a potential critical transition. The less dominant eigenvectors provide a lens to examine such situations, but can also undergo critical transitions themselves. 

A social group centrality model is developed and analysed to help study and explain the critical transitions in centrality. The degree of nodes and the popularity distribution of their neighbours play an important role in the changing centrality. This model presents circumstances in which critical transitions occurs and connects this to changing degrees. 

Future work will extend this analysis with applications to other networks where both connection and meta-information, such as popularity, are available, providing more exploration to further explain these critical transitions in large graphs.

\section{Availability of data and materials}
	The full data accessed from Spotify including both the network and metadata, as well as the code to generate the SGC model and analyse both networks, will be made publicly available upon acceptance.


 \section{Funding}
   The authors received no specific funding for this work.

\section{Author's contributions}
  TS performed data collection, analysis, model implementation and visualisation. MR identified critical transition behaviour and eigenvalue explanation. LM provided result interpretation and supervision. All authors co-wrote, read and approved the final manuscript. 

\section{Acknowledgements}
TS would like to acknowledge the support of the Australian Mathematical Sciences Institute (AMSI)  through it's Vacation Research Scholarships, from which the idea for this research was catalysed.

\bibliographystyle{comnet}
\bibliography{spotify}    

\begin{thebibliography}{00}

\bibitem{noauthor_spotify_nodate}
 {{Spotify}} \textemdash{} {{Company Info}}.
  https://newsroom.spotify.com/company-info/.

\bibitem{albert_statistical_2002}
Albert, R. {\&} Barab{\'a}si, A.-L. (2002)  Statistical Mechanics of Complex
  Networks. {\em Reviews of modern physics}, \textbf{74}(1), 47.

\bibitem{bae_multi-scale_2016}
Bae, A., Park, D., Ahn, Y.-Y. {\&} Park, J. (2016)  The {{Multi}}-{{Scale
  Network Landscape}} of {{Collaboration}}. {\em PLOS ONE}, \textbf{11}(3),
  e0151784.

\bibitem{barabasi_emergence_1999}
Barab{\'a}si, A.-L. {\&} Albert, R. (1999)  Emergence of {{Scaling}} in
  {{Random Networks}}. {\em Science}, \textbf{286}(5439), 509--512.

\bibitem{bonacich_factoring_1972}
Bonacich, P. (1972)  Factoring and Weighting Approaches to Status Scores and
  Clique Identification. {\em The Journal of Mathematical Sociology},
  \textbf{2}(1), 113--120.

\bibitem{bonacich_power_1987}
Bonacich, P. (1987)  Power and {{Centrality}}: {{A Family}} of {{Measures}}.
  {\em American Journal of Sociology}, \textbf{92}(5), 1170--1182.

\bibitem{borgatti_robustness_2006}
Borgatti, S.~P., Carley, K.~M. {\&} Krackhardt, D. (2006)  On the Robustness of
  Centrality Measures under Conditions of Imperfect Data. {\em Social
  Networks}, \textbf{28}(2), 124--136.

\bibitem{costa_characterization_2007}
Costa, L. d.~F., Rodrigues, F.~A., Travieso, G. {\&} Villas~Boas, P.~R. (2007)
  Characterization of Complex Networks: {{A}} Survey of Measurements. {\em
  Advances in Physics}, \textbf{56}(1), 167--242.

\bibitem{costenbader_stability_2003}
Costenbader, E. {\&} Valente, T.~W. (2003)  The Stability of Centrality
  Measures When Networks Are Sampled. {\em Social Networks}, \textbf{25}(4),
  283--307.

\bibitem{crossley_pretty_2008}
Crossley, N. (2008)  Pretty {{Connected}}: {{The Social Network}} of the
  {{Early UK Punk Movement}}. {\em Theory, Culture \& Society}, \textbf{25}(6),
  89--116.

\bibitem{cvetkovic_spectra_1980}
Cvetkovic, D.~M., Doob, M. {\&} Sachs, H. (1980) {\em Spectra of Graphs},
  volume~10.
{Academic Press, New York}.

\bibitem{dimaggio_social_2001}
DiMaggio, P., Hargittai, E., Neuman, W.~R. {\&} Robinson, J.~P. (2001)  Social
  {{Implications}} of the {{Internet}}. {\em Annual Review of Sociology},
  \textbf{27}(1), 307--336.

\bibitem{garg_measuring_2011}
Garg, R., Smith, M.~D. {\&} Telang, R. (2011)  Measuring {{Information
  Diffusion}} in an {{Online Community}}. {\em Journal of Management
  Information Systems}, \textbf{28}(2), 11--38.

\bibitem{gleiser_community_2003}
Gleiser, P.~M. {\&} Danon, L. (2003)  Community Structure in Jazz. {\em
  Advances in Complex Systems}, \textbf{06}(04), 565--573.

\bibitem{haveliwala_second_2003}
Haveliwala, T. {\&} Kamvar, S. (2003)  The Second Eigenvalue of the {{Google}}
  Matrix. Technical report, {Stanford}.

\bibitem{janosov_elites_2020}
Janosov, M., Musciotto, F., Battiston, F. {\&} I{\~n}iguez, G. (2020)  Elites,
  Communities and the Limited Benefits of Mentorship in Electronic Music. {\em
  Scientific Reports}, \textbf{10}(1), 3136.

\bibitem{klimek_fashion_2019}
Klimek, P., Kreuzbauer, R. {\&} Thurner, S. (2019)  Fashion and Art Cycles Are
  Driven by Counter-Dominance Signals of Elite Competition: Quantitative
  Evidence from Music Styles. {\em Journal of The Royal Society Interface},
  \textbf{16}(151), 20180731.

\bibitem{kruse_local_2010}
Kruse, H. (2010)  Local {{Identity}} and {{Independent Music Scenes}},
  {{Online}} and {{Off}}. {\em Popular Music and Society}, \textbf{33}(5),
  625--639.

\bibitem{martin_localization_2014}
Martin, T., Zhang, X. {\&} Newman, M. E.~J. (2014)  Localization and Centrality
  in Networks. {\em Physical Review E}, \textbf{90}(5), 052808.

\bibitem{mcandrew_music_2015}
McAndrew, S. {\&} Everett, M. (2015)  Music as {{Collective Invention}}: {{A
  Social Network Analysis}} of {{Composers}}. {\em Cultural Sociology},
  \textbf{9}(1), 56--80.

\bibitem{newman_mixing_2003}
Newman, M. E.~J. (2003)  Mixing Patterns in Networks. {\em Physical Review E},
  \textbf{67}(2), 026126.

\bibitem{newman_mathematics_2008}
Newman, M. E.~J. (2008)  The {{Mathematics}} of {{Networks}}. In {\em The {{New
  Palgrave Dictionary}} of {{Economics}}}, pages 8525--8533. {Palgrave
  Macmillan UK}, second edition.

\bibitem{newman_networks_2018}
Newman, M. E.~J. (2018) {\em Networks}.
{Oxford University Press}, second edition.

\bibitem{ng_link_2001}
Ng, A.~Y., Zheng, A.~X. {\&} Jordan, M.~I. (2001a)  Link Analysis, Eigenvectors
  and Stability. In {\em International {{Joint Conference}} on {{Artificial
  Intelligence}}}, volume~17, pages 903--910. {Lawrence Erlbaum Associates
  Ltd}.

\bibitem{ng_stable_2001}
Ng, A.~Y., Zheng, A.~X. {\&} Jordan, M.~I. (2001b)  Stable Algorithms for Link
  Analysis. In {\em Proceedings of the 24th Annual International {{ACM SIGIR}}
  Conference on {{Research}} and Development in Information Retrieval}, pages
  258--266.

\bibitem{niu_robustness_2015}
Niu, Q., Zeng, A., Fan, Y. {\&} Di, Z. (2015)  Robustness of Centrality
  Measures against Network Manipulation. {\em Physica A: Statistical Mechanics
  and its Applications}, \textbf{438}, 124--131.

\bibitem{page_pagerank_1999}
Page, L., Brin, S., Motwani, R. {\&} Winograd, T. (1999)  The PageRank Citation
  Ranking: Bringing Order to the Web.. Technical report, Stanford InfoLab.

\bibitem{park_topology_2015}
Park, D., Bae, A., Schich, M. {\&} Park, J. (2015)  Topology and Evolution of
  the Network of Western Classical Music Composers. {\em EPJ Data Science},
  \textbf{4}(1), 1--15.

\bibitem{park_social_2007}
Park, J., Celma, O., Koppenberger, M., Cano, P. {\&} Buld{\'u}, J.~M. (2007)
  The Social Network of Contemporary Popular Musicians. {\em International
  Journal of Bifurcation and Chaos}, \textbf{17}(07), 2281--2288.

\bibitem{perc_beauty_2020}
Perc, M. (2020)  Beauty in Artistic Expressions through the Eyes of Networks
  and Physics. {\em Journal of The Royal Society Interface}, \textbf{17}(164),
  20190686.

\bibitem{rodrigues_network_2019}
Rodrigues, F.~A. (2019)  Network Centrality: An Introduction. In {\em A
  Mathematical Modeling Approach from Nonlinear Dynamics to Complex Systems},
  pages 177--196. {Springer}.

\bibitem{sigaki_history_2018}
Sigaki, H. Y.~D., Perc, M. {\&} Ribeiro, H.~V. (2018)  History of Art Paintings
  through the Lens of Entropy and Complexity. {\em Proceedings of the National
  Academy of Sciences}, \textbf{115}(37), E8585--E8594.

\bibitem{wasserman_social_1994}
Wasserman, S. {\&} Faust, K. (1994) {\em Social {{Network Analysis}}:
  {{Methods}} and {{Applications}}}.
{Cambridge University Press}.

\bibitem{youngblood_cultural_2019}
Youngblood, M. (2019)  Cultural Transmission Modes of Music Sampling Traditions
  Remain Stable Despite Delocalization in the Digital Age. {\em PLOS ONE},
  \textbf{14}(2), e0211860.

\end{thebibliography}

\end{document}